\documentclass[acmtog,nonacm]{acmart}

\usepackage{graphicx}
\usepackage{subcaption}

\usepackage{booktabs}

\usepackage{pgfplots}
\pgfplotsset{width=10cm,compat=1.8}
\usepackage{pgfplotstable}

\usepackage{multirow}
\usepackage{xpatch}
\usepackage{placeins}
\usepackage{caption}
\usepackage{amsmath}
\usepackage{makecell}

\usepackage{xcolor}

\usepackage[ruled]{algorithm2e} 

\SetAlFnt{\small}
\SetAlCapFnt{\small}
\SetAlCapNameFnt{\small}
\SetAlCapHSkip{0pt}

\let\oldnl\nl
\newcommand{\nonl}{\renewcommand{\nl}{\let\nl\oldnl}}

\newlength\inlen
\newcommand\myinput[1]{%
  \nonl\settowidth\inlen{\KwIn{}}%
  \setlength\hangindent{\inlen}%
  \hspace*{\inlen}#1\\}

\newlength\outlen

\usepackage{mathtools}

\def \bezier {B{\'e}zier}

\newcommand{\Gtwo}{G_{\text{2D}}}
\newcommand{\Gthree}{G_{\text{3D}}}
\newcommand{\R}{\mathbb{R}}

\begin{document}
\title{Spatially Accelerated Winding Numbers for Curved Geometry}

\author{Jacob Spainhour}
\orcid{0000-0001-8219-4360}
\affiliation{%
  \institution{Lawrence Livermore National Laboratory}
  \city{Livermore}
  \country{USA}}
\email{jspainhour@llnl.gov}

\author{Brad Whitlock}
\orcid{0000-0001-6261-7522}
\affiliation{%
  \institution{Lawrence Livermore National Laboratory}
  \city{Livermore}
  \country{USA}}
\email{whitlock2@llnl.gov}

\author{Kenneth Weiss}
\orcid{0000-0001-6649-8022}
\affiliation{%
  \institution{Lawrence Livermore National Laboratory}
  \city{Livermore}
  \country{USA}}
\email{kweiss@llnl.gov}

\renewcommand\shortauthors{J. Spainhour, B. Whitlock, and K. Weiss}

\begin{abstract}
The generalized winding number (GWN) is a scalar field that supports robust containment queries on curved geometry, including non-watertight, overlapping, and nested boundary representations.
While queries can be easily parallelized over samples, direct evaluation on parametric curves and surfaces remains costly for large and complex models.
Fast, state-of-the-art GWN approaches leverage a spatial index to approximate the GWN, typically coupled with a Taylor expansion which approximates the GWN contribution for far clusters of geometric primitives. 
However, such methods operate only on discrete inputs such as triangle meshes and point clouds, and would introduce containment errors near boundaries if applied to curved input.
We extend support for fast GWN evaluation over arbitrary collections of NURBS curves in 2D and trimmed NURBS patches in 3D via a Bounding Volume Hierarchy that stores efficiently precomputed moment data in the hierarchy nodes.
When querying the hierarchy, approximations for far clusters are used alongside direct evaluation for nearby NURBS primitives, achieving sub-linear complexity while preserving the geometric features in the vicinity of the query point.
Central to our performance improvements is an adaptive subdivision strategy for NURBS primitives during a preprocessing phase, creating better spatial partitions while retaining the same accuracy for containment decisions as a direct evaluation.
We demonstrate the performance and accuracy of our approach across a large collection of 2D and 3D datasets.

\end{abstract}

%
%

%
%


\begin{teaserfigure}
  \includegraphics[width=\linewidth]{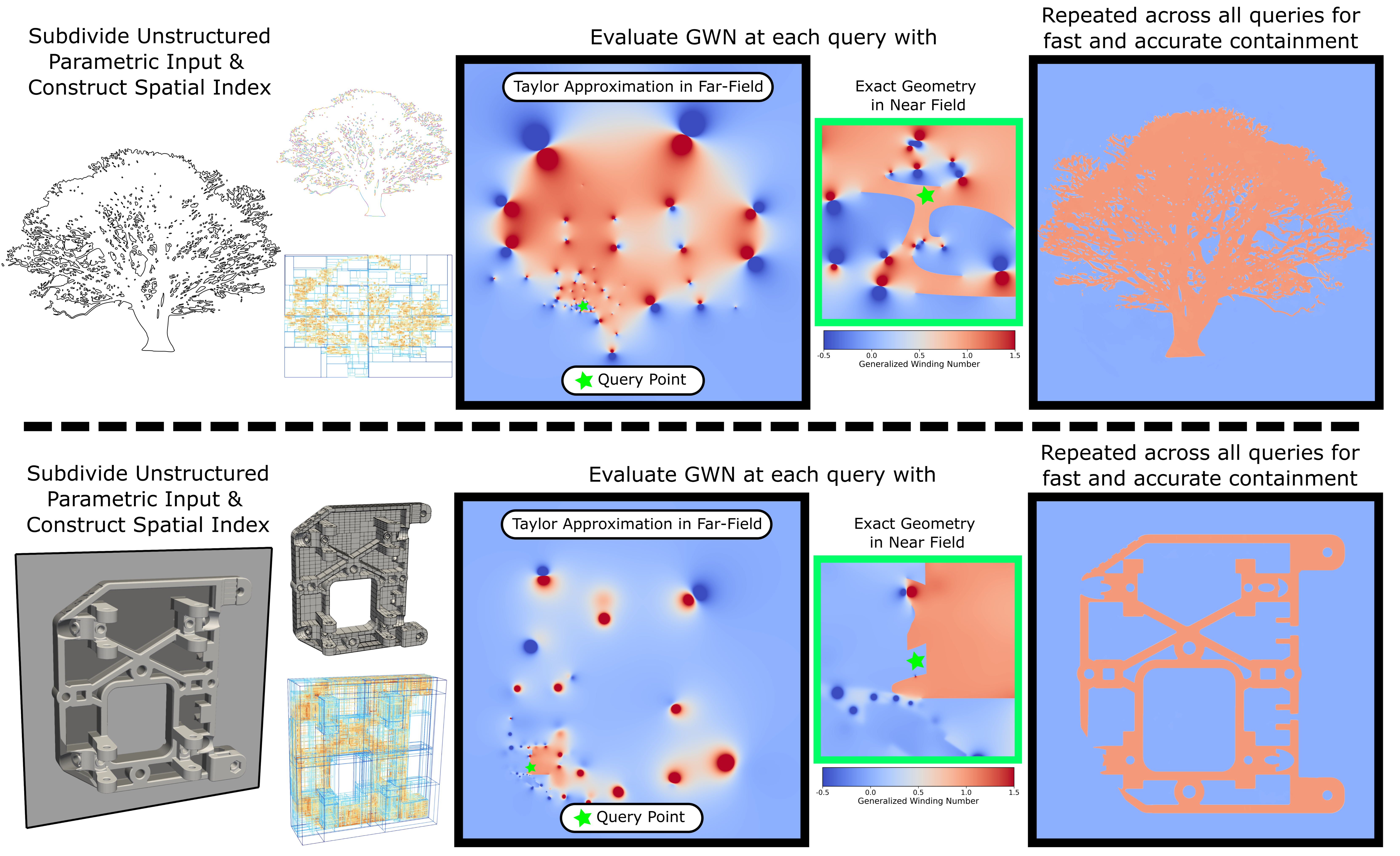}
  \caption{The Generalized Winding Number (GWN) facilitates containment queries for non-watertight shapes by treating each piece of the boundary representation (B-Rep) independently, but direct evaluation necessarily results in linear complexity.
  We address this by extending approximation strategies for the GWN for clusters of discrete geometric primitives to more general NURBS primitives in 2D and 3D, which are then leveraged in a spatial index which hierarchically clusters curved input.
  As preprocessing, we first adaptively subdivide the boundary to ensure efficient utilization of the spatial index, and calculate query-independent moment data which parameterizes the GWN approximation.
  Given a query point, the spatial index is traversed to identify clusters which are sufficiently far to be well-approximated.
  Otherwise, nearby B-Rep components are treated with exact GWN evaluation methods, thereby ensuring fast evaluation of containment queries that respect all curved features of the input at all points in space.
  }
  \Description[Graphical Abstract]{The proposed agglomeration pipeline is demonstrated for a 2D SVG example composed of NUBRS curves and a 3D CAD example composed of trimmed NURBS surfaces. For each shape, the original shape is shown, its subdivided NUBRS primitives, and the generated BVH tree. The scalar field which is used to compute the GWN at a specific query point is shown. Far away from the point, the field is represented as dipoles representing clusters of boundary objects. Shown on a zoomed-in area around the query point is the exact GWN field of the nearest boundary components. For each shape, the fully approximated GWN field is shown, which is accurate across the domain.}
	\label{fig:graphical_abstract}
\end{teaserfigure}

\maketitle

\section{Introduction}\label{sec:introduction}
Containment queries are staple geometric operations whose practical evaluation is often complicated by the presence of non-watertight geometry~\cite{sederberg-08-watertightnurbs,oropallo-slicing-2017,xiao-2021-nurbsrepair,weiss-16-shaping}.
In light of this, there has been a recent focus in the literature on evaluation methods for the generalized winding number (GWN), a mathematical construct which defines a reasonable proxy for containment of query points within shapes whose boundary does not define an interior volume.

Intuitively, the GWN at a query point is the fraction of its field of view which is occupied by the given boundary representation (B-Rep), equivalent to a signed angle in 2D and a signed solid angle in 3D.
Just as gaps and overlaps between B-Rep components become less apparent when viewed from further away, the influence of geometric discontinuities on the GWN field decay rapidly.
Conversely, the most important features of a GWN containment method are how it treats nearby boundary components in order to remain faithful to continuous portions of the shape.

However, the numerical value of the GWN itself is a \textit{global} property of the entire shape.
Even if the individual contributions are small, the sum across B-Rep components must still be recorded with enough precision for the subsequent containment decision to be trusted.
This represents a fundamental limitation on the evaluation of GWN methods, as the cost of exact evaluation scales linearly with the number of boundary components.

To overcome this, we develop what we refer to as an \textit{agglomeration} strategy, which reduces algorithm complexity to sub-linear by identifying clusters of B-Rep components and approximating the GWN with respect to the entire collection with simple formulae.
This naming is meant to suggest a loose grouping of B-Rep components, as they are only treated as part of a cluster when it is known that the resulting approximation error is small.
It also contrasts with the idea of a stricter \textit{aggregation} strategy, which utilizes topological information from the input mesh data structure (as in~\citet{Jacobson-13-winding} or~\citet{bao-2025-ellipsemethod}), or could apply a global, tolerance-dependent welding of adjacent B-Rep components, in each case accelerating performance by avoiding redundant calculations.

Existing agglomeration strategies have been applied only to discrete geometric primitives, e.g., point clouds and shapes composed of linear elements.
In this work, we seek to extend these methods to input collections of NURBS curves in 2D and trimmed NURBS surfaces in 3D.
We present the following principal contributions:
\begin{itemize}
    \item We synthesize spatial acceleration strategies~\cite{Barill-18-soupcloud} and curve-preserving GWN methods~\cite{spainhour_24_robustcontainment2d,spainhour_25_robustcontainment3d} into a pair of fast GWN methods which are capable of producing accurate containment decisions at arbitrary points at its default settings with sub-linear runtime complexity.
    \item We describe our method for constructing a Bounding Volume Hierarchy (BVH) of curved objects for the purpose of fast GWN evaluation.
    This involves efficient evaluation of geometric data for clusters of curved primitives, and an adaptive subdivision strategy for parametric input to produce a more balanced BVH. 
    Both are critical to extracting performance from the spatial index in the context of curved geometry.
    \item We improve the state-of-the-art performance of direct GWN evaluation methods for NURBS objects, which we utilize in our algorithm's processing of leaf-nodes. We have also developed an open-source benchmark for evaluating and comparing 2D direct GWN methods on the same computing platform using shared infrastructure.
\end{itemize}

\section{Background and Related Work}\label{sec:background}
\begin{figure*}[t]
  \includegraphics[width=1.0\linewidth]{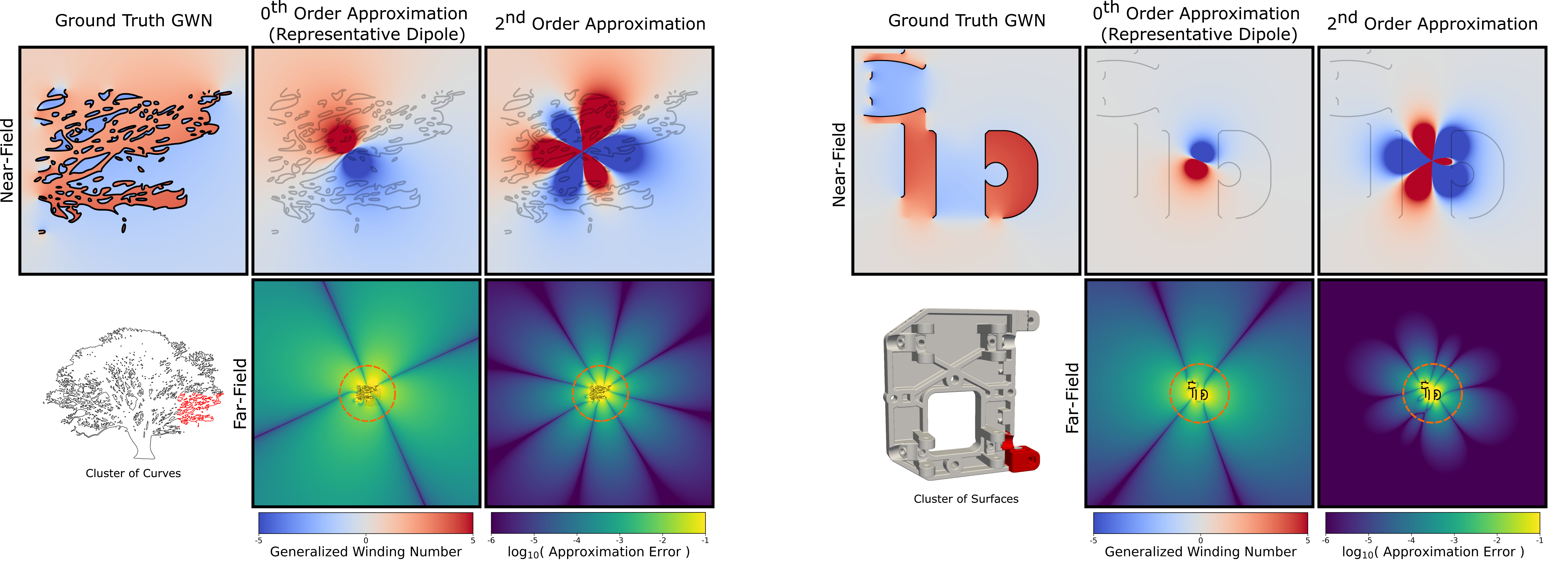}
    \caption{We show the $0^\text{th}$ order and $2^\text{nd}$ order Taylor expansions utilized by our agglomeration strategy to represent the GWN field of a single cluster of 2D curves and 3D NURBS trimmed surfaces, with each approximation centered at the centroid of the components shapes.
    We view the approximation near the boundaries (top), and the error in their far field (bottom). 
    On these error plots, we also show the radius at which our proposed algorithm considers query points to be ``sufficiently far'' for the approximation to be reasonably utilized ($\beta = 2$).
  }
  \Description[Illustration of Taylor expansion for clusters]{For a 2D and 3D example, the Taylor expansion of orders 0 and 2 is shown. The order 0 expansion is a dipole matching the orientation of the ground-truth GWN field, and the order 2 expansion has a similar orientation, but extra lobes on the dipole. On a zoomed out view around the cluster, the absolute error in the Taylor approximation is shown, and while the error has similar lobes as the approximation itself, the error decays away from the query.}
  \label{fig:agglomeration_demo}
\end{figure*}

A straightforward approach to improving GWN evaluation time is simply increasing the efficiency with which individual B-Rep components are processed.
For example, recent performance improvements to 2D GWN evaluation focus on the nearest boundary curves, for which the GWN is necessarily the most expensive to compute.
Current approaches include geometric clipping algorithms~\cite{spainhour_24_robustcontainment2d,bao-2025-ellipsemethod}, algebraic root-finding methods~\cite{liu-2025-closedform-wn}, or methods which leverage structured collections of input points for more efficient ray-casting~\cite{martens-2025-oneshot}.
However, there are fundamentally fewer improvements available for better handling individual far-away curves, as evaluation is optimized by direct evaluation of the curve's signed angle through a single, and as of yet unavoidable trigonometric operation~\cite{spainhour_24_robustcontainment2d}.

This trend is seen among 3D methods as well, as most concern themselves with accurate processing of nearby curved surfaces~\cite{sawhney-20-montecarlogeometry,spainhour_25_robustcontainment3d,martens-2025-oneshot}, although the recency of these methods leaves more potential for improvement for far-field evaluation.

In either case, even if far-away objects can be processed at less individual expense, their total cost inevitably dominates as the number of B-Rep components in a shape increases.
Indeed, the GWN was first conceptualized in~\citet{Jacobson-13-winding} to robustly determine containment in large non-watertight triangle meshes for which a naive triangle-by-triangle evaluation would be unnecessarily costly.
Instead, that work evaluated the GWN for local triangle collections with an exact evaluation method that scaled with the number of exterior edges rather than constituent triangle boundaries.
This was paired with a divide-and-conquer algorithm which constructs a BVH from subdivisions of connected components, achieving sub-linear complexity.
While that method produces an exact GWN value at every input point, its asymptotic performance relies on the assumption that the input mesh is reasonably well-tessellated, with as few connected components as possible. 

However, many applications can make no such structural assumptions on the arrangement nor connectivity of the input boundary components.
For example, the method of~\citet{Barill-18-soupcloud} defines the GWN for unstructured point cloud data as the area-weighted sum of dipole contributions.
In that work, logarithmic complexity is achieved for such input by approximating the GWN for clusters of dipoles with a scalar field that can be rapidly evaluated via a Taylor expansion around the cluster center.
Intuitively, a zeroth order Taylor expansion is equivalent to the scalar field generated by a single, more heavily weighted dipole whose orientation is determined by the weighted average of its constituent dipoles.
That work also extends this formulation to triangle soups, where the parameters of the Taylor expansion are summed across the primitives in each cluster as preprocessing.

A key feature of each of these methods is that they both produce a containment decision that is \textit{accurate} with respect to the input shape, in the sense that it is essentially impossible for either method to produce an off-by-one error that would result in a containment misclassification.
This is true despite producing a GWN value that is less \textit{precise} than direct evaluation, i.e., subject to error in its fractional component.
This is because approximations are used only in cases where the error is known to be insufficient to disturb the subsequent containment decision.
Otherwise, the GWN is computed directly from the provided boundary without approximation. 

Using one of these agglomeration methods with general curved geometry (e.g., curves in 2D and surfaces in 3D) would require first discretizing the geometry into a linearization/triangulation, or sampling from it a point cloud~\cite{balu-23-immersogeometric}.
This means that even if the containment decision is accurate with respect to the discretized geometry, it will almost certainly result in misclassifications for points near to the \textit{original} curved boundary.

In this work, we develop an agglomeration strategy akin to~\citet{Barill-18-soupcloud} that is specialized to curved primitives, namely NURBS curves in 2D and trimmed NURBS surfaces in 3D. 
The clear advantage of such an approach is that we retain the fast, error-controlled approximation of the GWN for entire clusters of B-Rep components far from the query point.
We visualize such an approximation in Figure~\ref{fig:agglomeration_demo}, observing how the orientation of the approximation corresponds to the ground-truth GWN field, and how the quality of the approximation improves rapidly away from the cluster centroid.
Furthermore, utilizing near-field evaluation methods which exactly respect the curved features of the input provides near-guarantees for the accuracy of the containment decision~\cite{spainhour_24_robustcontainment2d,spainhour_25_robustcontainment3d}. 

There are also advantages to directly using the parametric formulation of the shape boundary.
Compared to an approach which evaluates the GWN from a point cloud sampled from the shape boundary, computing the Taylor expansion coefficients directly from the shape boundary via numerical integration necessarily improves precision in the computed field.
More significantly, having an exact parametric representation of the boundary adds a considerable amount of flexibility to the total number of clusters considered by our method, as we can artificially create new B-Rep components by subdividing those which are provided.
We discuss the specific advantages of doing so in Section~\ref{sec:subdivision}.

\subsection{Taylor Expansion for Curved Geometry}\label{sec:formalism}
Mathematically, the generalized winding number $w$ at a query point $q$ with respect to a collection of boundary elements $\Gamma_i$ (curves $\{C_i\}$ in 2D or surfaces $\{S_i\}$ in 3D) is given by the sum of the (solid) angles spanned by each component~\cite{Jacobson-13-winding}:
\begin{align}
    \label{eqn:gwn_sum}
    w_C(q) := \sum_i \int_{C_i} \,d\theta\quad\quad\quad w_S(q) := \sum_i \int_{S_i} \,d\Omega
\end{align}

Notably, each formulation in Cartesian coordinates is equivalent to an integral over the normal derivative of the relevant Green's function $G(x; q)$ of the Laplace equation~\cite{evans2010partial}.
Given an arbitrary boundary component $\Gamma$ in $\R^2$ or $\R^3$, we have
\begin{align}
    \label{eqn:gwn_greens}
    w_\Gamma(q) = \int_\Gamma \widehat{n}\cdot \nabla G(x; q)\,dx, \quad\begin{cases}
        \Gtwo(x; q) = \frac{1}{2\pi}\ln||x - q||\\
        \Gthree(x; q) = -\frac{1}{4\pi}||x - q||^{-1}
    \end{cases}
\end{align}

Following~\citet{Barill-18-soupcloud}, we can combine this formulation with a Taylor expansion of the normal derivative $\widehat{n}\cdot \nabla G(x; q)$ centered at point $x_0$ to produce an approximation for the GWN field generated by the boundary components $\{\Gamma_i\}$: 
\begin{align}
\label{eqn:taylor}
    w_\Gamma(q) \approx &\sum_{\Gamma_i\in\Gamma} \left(\int_{\Gamma_i} \widehat{n}\,dx \right) \cdot \nabla G(x_0; q) \\
                                      +&\sum_{\Gamma_i\in\Gamma} \left(\int_{\Gamma_i} (x - x_0)\otimes\widehat{n}\,dx \right)\cdot \nabla^2 G(x_0; q) \nonumber\\
                                      +&\tfrac{1}{2}\sum_{\Gamma_i\in\Gamma} \left(\int_{\Gamma_i} (x - x_0)\otimes(x - x_0)\otimes \widehat{n}\,dx \right) \cdot \nabla^3G(x_0; q). \nonumber
\end{align}
where $\nabla G(\cdot; q) : \R^d \to \R^d$, $\nabla^2 G(\cdot; q): \R^d \to \R^{d^2}$, and $\nabla^3 G(\cdot; q) : \R^d \to \R^{d^3}$ (See~\citet{Barill-18-soupcloud} for an expression of each term.)

Importantly, only the evaluation of the relevant derivative of $G$ depends on the query point $q$. 
In contrast, the coefficients for each term in the expansion (each the sum of integrals over each B-rep component) can be precomputed and stored across clusters, which are then leveraged in a hierarchical approximation of the GWN for an entire shape.

\section{Method}\label{sec:methods}
Briefly, our proposed method first collects all curved boundaries into a BVH according to the axis-aligned bounding box (AABB) of individual B-Rep components.
Within the BVH, each leaf node represents a single NURBS curve in 2D or trimmed NURBS surface in 3D,
while each internal node represents a candidate cluster whose GWN field is approximated by the Taylor expansion described in~\citet{Barill-18-soupcloud} and reiterated in Section~\ref{sec:formalism}. 
This Taylor expansion is parameterized by geometric moments computed across all leaf nodes during BVH initialization, described in Section~\ref{sec:initialization}. 

By traversing the BVH at a given query point, we identify the largest clusters which are far from the query.
Following~\citet{Barill-18-soupcloud}, we define a ``far-away'' cluster as one for which the query point is outside a sphere whose center is the centroid of the constituent B-Rep components, and whose radius is larger than the AABB radius by a factor of $\beta$ (we take $\beta=2$ by default).
We also identify the leaf nodes which contain the query, for which we evaluate the GWN of their NURBS primitives with the method of~\citet{spainhour_24_robustcontainment2d} in 2D and~\citet{spainhour_25_robustcontainment3d} in 3D.
These two subroutines have been designed to maintain the exact geometric fidelity of the input geometry, thereby ensuring that correct containment decisions are achieved for points at arbitrary locations, even very near to the surface.
We discuss our implementation of each method in Section~\ref{sec:direct_eval}.

The complete algorithm is presented in Alg.~\ref{alg:agglomerated_gwn}, and has been implemented in Axom, a BSD-licensed open-source library of computer science infrastructure for HPC applications~\cite{Axom_CS_infrastructure}.

\subsection{BVH Initialization}\label{sec:initialization}
We use a Karras-style BVH binary tree~\cite{karras2012maximizing} to generate a spatial index over our curved boundary components.
Given a cluster of curved boundary components, it is necessary to accurately compute the geometric data which parameterizes the Taylor expansion by evaluating moment integrals of the forms in Equation~\ref{eqn:taylor} as well as the cluster centroids.
For general curves and curved surfaces, these integrals have no exact analytic solution, and must be evaluated numerically, such as with quadrature.
Furthermore, these integrals (for orders higher than 0) must be evaluated with respect to the cluster center $x_0$, with the number of clusters generally scaling quadratically with the number of input primitives.
This means that direct evaluation for each cluster would make the preprocessing time impractical.

However, each of these ``centered'' moments is related to an ``uncentered'' counterpart that can be evaluated independently of the specific center.
For example, we can evaluate first order moments through the transform
\begin{align}\label{eqn:moment_transform}
    \int (x - x_0)\otimes\widehat{n}\,dx = \int x\otimes\widehat{n}\,dx - x_0 \otimes \int \widehat{n}\,dx,
\end{align}
with an analogous formula existing for the $2^{nd}$ order parameters.

This means that during preprocessing, we only need to evaluate the uncentered moments given by
\begin{align}
    \int_{\Gamma_i} \widehat{n}\,dx,\quad\quad\int_{\Gamma_i} x\otimes\widehat{n}\,dx,\quad\quad\int_{\Gamma_i} x\otimes x\otimes\widehat{n}\,dx,
\end{align}
giving a total of $(2 + 4 + 8)$ values in 2D and $(3 + 9 + 27)$ values in 3D for each leaf node.

Importantly, uncentered moments for a given internal node can be evaluated during BVH initialization by summing the analogous values of its children, at which point the uncentered moments for the children can be transformed in place to the actual quantities of interest.
All that remains is to evaluate these uncentered moments for each boundary component.

For 2D curves, we utilize a fact demonstrated by~\citet{spainhour_24_robustcontainment2d}: for any query point sufficiently far from 2D input curve $C$, the GWN with respect to $C$ is equal to the GWN of the line connecting its two endpoints.
This means that we can instead parameterize the Taylor expansion according to an analogous cluster of linear segments, as the query point is necessarily far from the cluster by construction. 
This results in a different approximation of the same field with the same Taylor expansion order, but one for which the requisite geometric data can be evaluated directly and exactly using standard formulae.

\begin{figure}[t]
    \centering
    \begin{minipage}[t]{0.22\linewidth}
        \centering
        \vspace{0pt}
        \includegraphics[width=\linewidth]{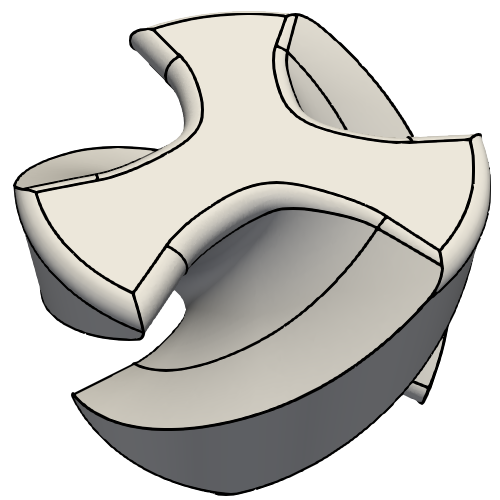}

        \vspace{0.5em}

        \includegraphics[width=\linewidth]{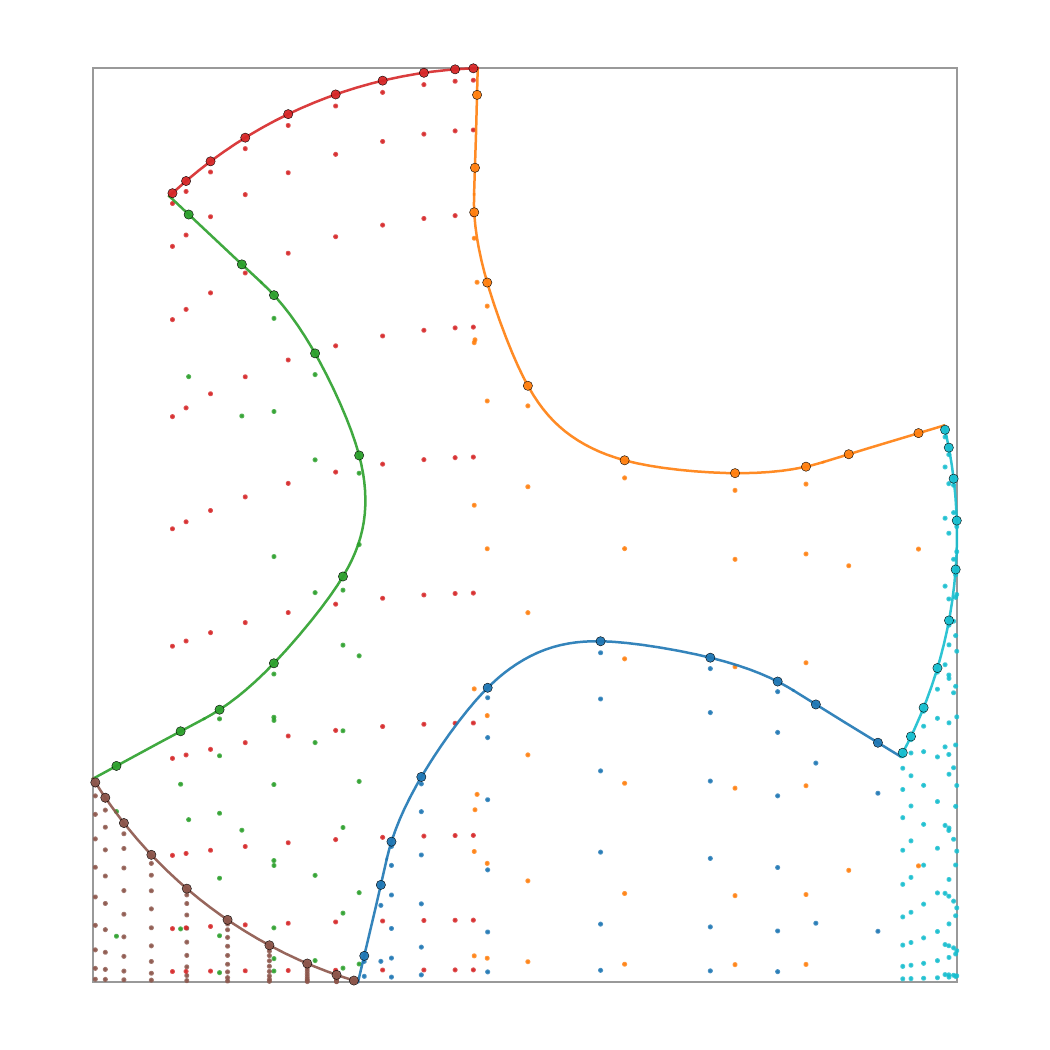}
    \end{minipage}
    \hfill
    \begin{minipage}[t]{0.77\linewidth}
        \centering
        \vspace{0pt}
        \includegraphics[width=\linewidth]{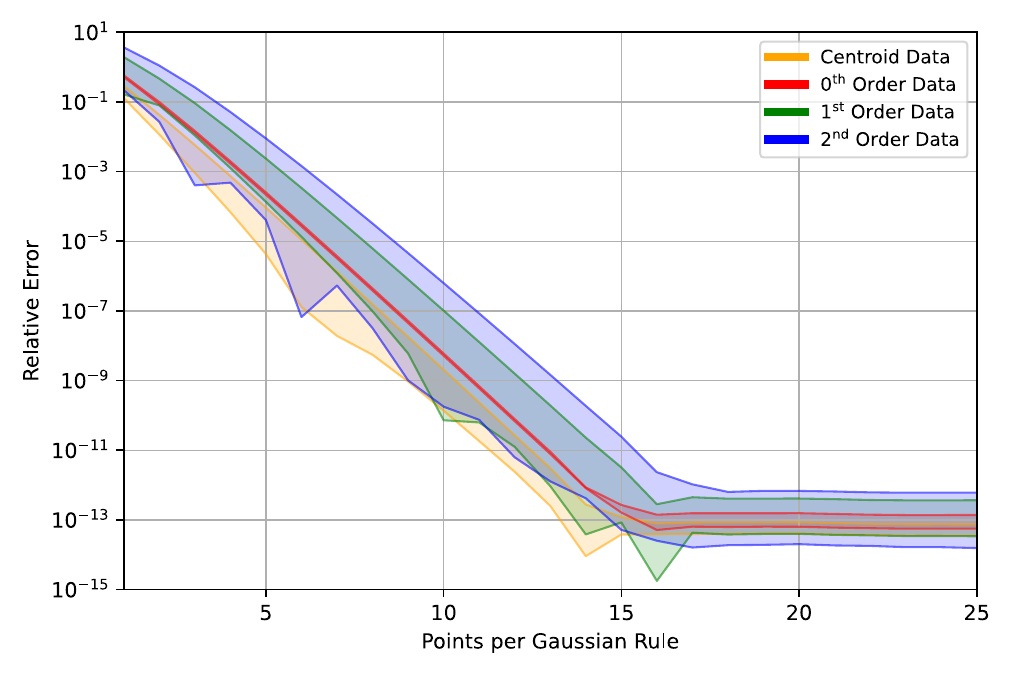}
    \end{minipage}

    \caption{We use numerical quadrature to efficiently and accurately evaluate 43 integrals encoding geometric data about each trimmed surface in a CAD model (4 for the centroid, and 3 + 9 + 27 across moments). In the lower-left subfigure, we illustrate how the nodes for each 10-point Gaussian rule are scattered in a patch's parameter space with respect to its trimming curves. The quadrature nodes can be reused across each integrated quantity. }
    \Description[Quadrature on a trimmed surface]{Three subfigures illustrate numerical quadrature for a trimmed NURBS surface. The left subfigure shows a rendering of a single trimmed surface patch. The middle-left subfigure shows the patch's 2D parameter space with multiple colored trimming curves and a scattered set of colored quadrature nodes. The right subfigure plots relative error versus points per Gaussian rule on a logarithmic y-axis, with separate curves for centroid data and 0th-, 1st-, and 2nd-order moment data.}
    \label{fig:quadrature_demo}
\end{figure}

In 3D, these integrals have an equivalent direct solution only when the surface is polygonal, and so we resort to numerical integration to handle the more general case of arbitrary trimmed NURBS.
Fortunately, the integrands for each quantity of interest do not exhibit the same near-singular behavior as the winding number itself in Equation~\ref{eqn:gwn_sum}, and so we can adapt existing quadrature methods.
We specifically use the quadrature approach of~\citet{gunderman-21-trimmednurbsintegration,gunderman-20-cad}, which evaluates surface integrals over trimmed NURBS surfaces in terms of area integrals over the surface trimming curves,
which are then evaluated using a numerical Green's theorem that scatters Gauss-Legendre quadrature nodes into the patch's parametric domain.
This method more faithfully represents the potentially complex boundary geometry of the surface than, say, a method which evaluates a tensor product over the untrimmed surface paired with an indicator function for its visible portions.

In the context of the proposed GWN method, we perform \bezier\ extraction during moment evaluation to produce continuous and smooth trimmed surfaces, and so the quadrature method exhibits exponential convergence.
This means that we achieve a high degree of precision in the moment calculation with relatively few nodes per Gaussian rule, but also that the quadrature is stable when more nodes are used than is strictly necessary to achieve the desired precision (see Figure~\ref{fig:quadrature_demo}).
As a result, we use $n = 10$ nodes per rule in the provided numerical results.
This is more than sufficient to ensure that the error in the resulting GWN evaluation is always dominated by the approximation error in the Taylor expansion, but does not incur unnecessary computational cost (See Section~\ref{sec:results_precision}).

\subsection{Subdivision of Curved Geometry}\label{sec:subdivision}
A specific advantage of working with NURBS input is that all curves and surfaces are defined explicitly through a parametric representation which can be manipulated without changing the shape.
Furthermore, it is possible for two shapes which are geometrically identical to result in significantly different average query times.
Within an agglomeration framework, for example, increasing the number of NURBS primitives also increases the amount of preprocessing required to compute moment data, as well as initialize and traverse the BVH.
However, it also typically decreases the time spent evaluating the GWN overall, as fewer query points lie in the AABB of a leaf node, and it is always faster to approximate the GWN through the Taylor expansion of a cluster than through direct evaluation.
At a certain point however, evaluation achieves diminishing returns while the increased preprocessing cost begins to dominate, especially as the extra storage requirements consume system memory.

Importantly, there is no simple ``sweet spot'' for the optimal number of primitives to target in the subdivision process.
Instead, we find that the overall spatial distribution of the bounding boxes of each leaf node is far more critical to efficient utilization of the BVH.
As such, we apply an adaptive subdivision routine to our NURBS input which limits the size of a shape's largest individual B-Rep components.
We bisect each 2D curve or 3D trimmed surface until its axis-aligned bounding-box has a diagonal with less than 10\% of the length of the diagonal for the AABB of the entire shape, a heuristic which we justify experimentally in Section~\ref{sec:subdivision_results}.
In 2D, this algorithm is preceded by \bezier\ extraction of all NURBS input, which is inexpensive for curves and largely satisfies our AABB size limit in of itself.
In 3D however, we find that the analogous extraction procedure often causes uneven and unhelpful refinements.

\begin{figure}[t]
  \includegraphics[width=\linewidth]{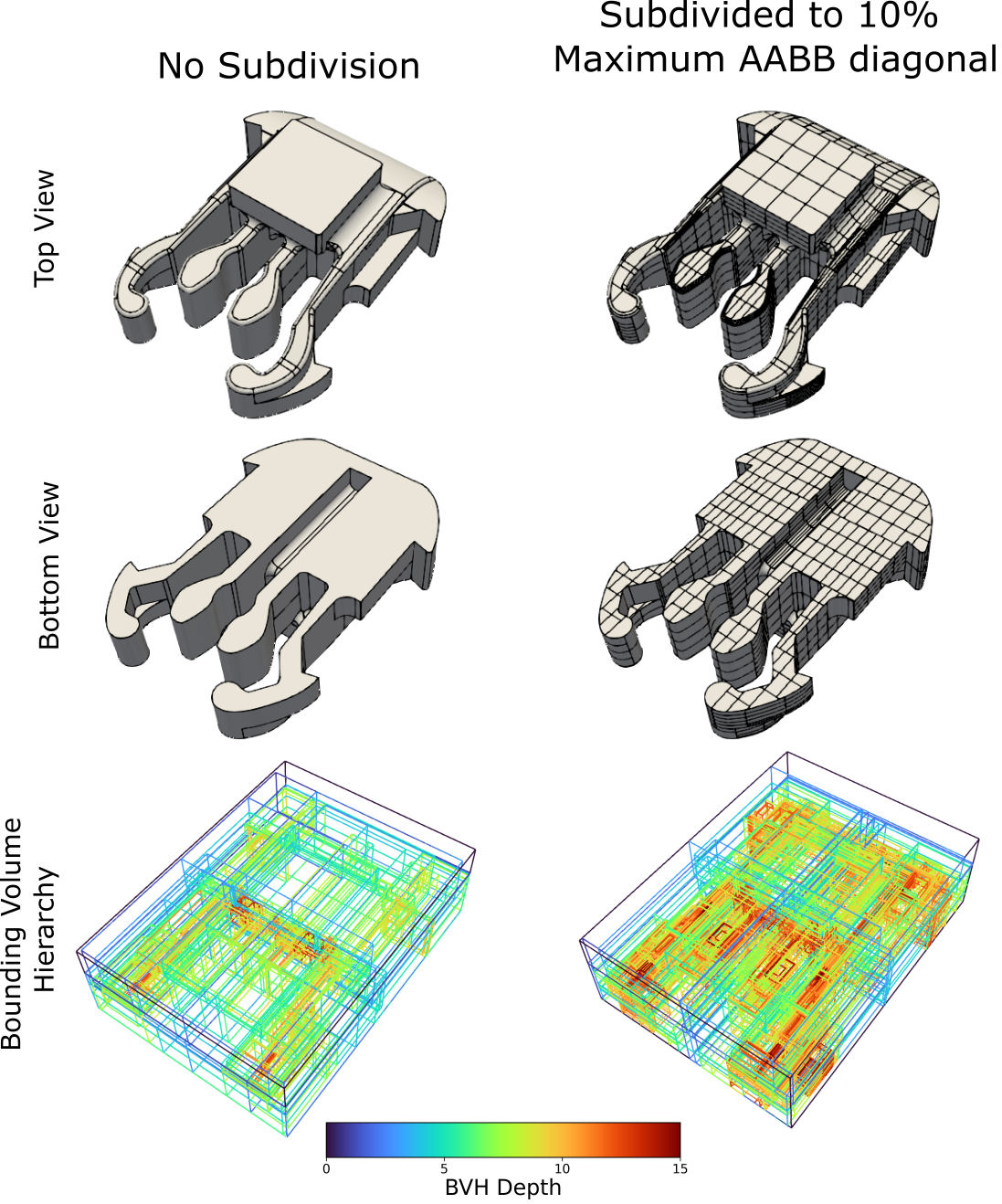}
  \caption{We demonstrate how subdividing parametric NURBS input influences the construction of the BVH, leading to a better spatial partition while leaving the underlying geometry unchanged.}
  \Description[Subdivision and BVH construction]{A 3D CAD model before and after adaptive subdivision is shown. The left column shows the model with no subdivision, and the right column shows the same model subdivided into a visible grid of smaller surface patches using a 10\% maximum AABB-diagonal threshold. For each case, two renderings are shown from different viewpoints, and a BVH visualization below shows many axis-aligned bounding boxes colored by tree depth.}
  \label{fig:subdivision_example}
\end{figure}

Intuitively, we can see in the example in Figure~\ref{fig:subdivision_example} that applying the 10\% bounding box refinement heuristic leads to a more balanced BVH with the same maximum depth.
In this example, the bottom trimmed surface has a bounding box which spans the entire shape, and so it would be evaluated as a leaf node in nearly all queries, severely limiting performance without subdivision.

\subsection{Direct Evaluation}\label{sec:direct_eval}

We base our direct GWN evaluation of curved primitives at BVH leaf nodes on the methods of~\citet{spainhour_24_robustcontainment2d} for 2D \bezier\ segments and~\citet{spainhour_25_robustcontainment3d} for 3D trimmed surfaces.
Because efficient direct evaluation dominates query times, we optimized our implementation beyond the methods in these works and their corresponding implementation in Axom~\cite{Axom_CS_infrastructure}.
For example, we improve the evaluation of the 2D GWN of linear components by computing subtended angles with an \texttt{atan2} call instead of \texttt{acos}, which avoids normalizing input vectors.
We similarly avoid all convex hull/point-in-polygon tests in favor of faster AABB containment checks.
We also improved on the memoization approach in 3D, i.e., dynamic caching and reusing per-trimming curve quadrature data, by optimizing and reducing memory allocations, and added similar functionality in 2D to cache and reuse curve subdivisions, bounding boxes, and convexity flags.
On comparable hardware, these changes yield a $2\text{--5}\times$ serial speedup compared to the reported timings in the respective manuscripts.

We also parallelized these routines with OpenMP threading.
While any GWN algorithm naturally parallelizes across query points, this required some additional care to make the caching structures supporting memoization thread safe.
In our implementation, each thread maintains a separate cache for the geometric objects,
and for evaluated quadrature nodes and tangents in 3D.
These changes have been incorporated into Axom's open source implementation.

\begin{algorithm}[tb]
    \caption{\texttt{AgglomeratedGWN} 
				Evaluate the generalized winding number for a collection of NURBS primitives. Note this algorithm applies to both 2D NURBS curves and 3D trimmed NURBS surfaces.
		}
    \SetCommentSty{textit}
    \DontPrintSemicolon
    \KwIn{$\{\Gamma\}$: Collection of NURBS objects}
    \myinput{$\{q\}$: Collection of query points}
    \KwOut{$\{w_{q}\}$: The GWN evaluated at each ${q}$}
    \nonl\;
		\tcp{Threaded preprocessing}
    $\{\Gamma\} \gets \texttt{AdaptiveSubdivision}(\{\Gamma\}, 10\%)$\;
    $\{M\} \gets \texttt{ComputeUncenteredMoments}(\{\Gamma\})$\;
    $\text{BVH} \gets \texttt{ConstructBVH}(\{\Gamma\})$\;
    $\texttt{SumUncenteredMomentsUpTree}(\text{BVH}, \{M\})$\;
    $\texttt{CenterMomentsInPlace}(\text{BVH}, \{M\})$\;
		\BlankLine
		\tcp{Threaded batched query}
    \ForEach{$q, w_q$ in $\{q\}, \{w_q\}$}{
        $w_q \gets 0$\;
				$S \gets \{\text{BVH.root}\}$\tcp*{Initialize stack with root}
        \While{$S$ not empty}{
            $\text{Node} \gets \texttt{Pop}(S)$\;
            \uIf{$\textnormal{Node}$ is a leaf}{
                $w_q \gets w_q + \texttt{DirectWindingNumber}(q, \text{Node.}\Gamma)$\;
            }\Else{
                \uIf{$||q - \textnormal{Node.centroid}|| > \beta\cdot\textnormal{Node.radius}$}{
                    $w_q \gets w_q + \texttt{ApproxWindingNumber}(q, \text{Node.}M)$\;
                }\Else{
                    $\texttt{Push}(S, \text{Node.children})$\;
                }
            }
        }
    }

    \Return $\{w_q\}$\;
    \label{alg:agglomerated_gwn}
\end{algorithm}

\section{Numerical Experiments and Results}
To efficiently evaluate our proposed algorithm, we construct a representative sample of 300 3D CAD models from the ABC-Dataset~\cite{koch-19-abcdataset} and a representative sample of 300 2D SVG models from OpenClipArt20k~\cite{openclipart}. 
These samples are generated from each collection by sampling across different orders of magnitude of shape complexity.
In 3D, we include 100 shapes stratified by patch count, with 25 drawn randomly from each of the bins [1, 10), [10, 100), [100, 1000), and $\geq$1000.
We also include 100 shapes stratified in the same way according to trimming curve count, and an additional 100 shapes sampled uniformly from the full collection.
In 2D, we select 150 shapes similarly stratified by curve count order of magnitude, and 150 sampled uniformly from the full collection.
All experiments were conducted with an Intel Xeon Platinum 8480+ (Sapphire Rapids) processor with 56 cores per socket and 256 GB of DDR5 memory.

\subsection{Adaptive Subdivision Algorithm}\label{sec:subdivision_results}
\begin{figure*}[t]
    \centering
    \begin{tabular}{cc}
        \includegraphics[width=0.48\linewidth]{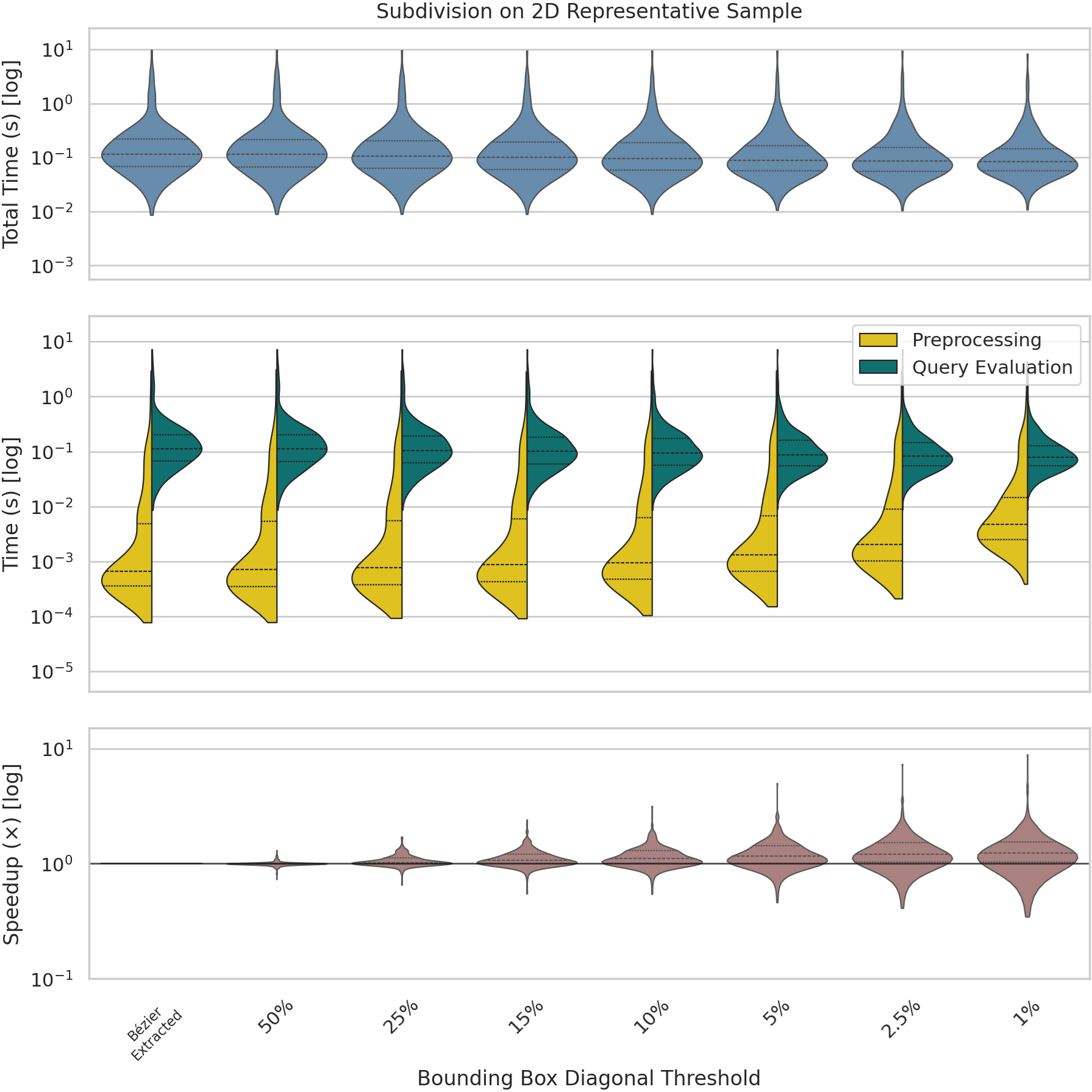}     & \includegraphics[width=0.48\linewidth]{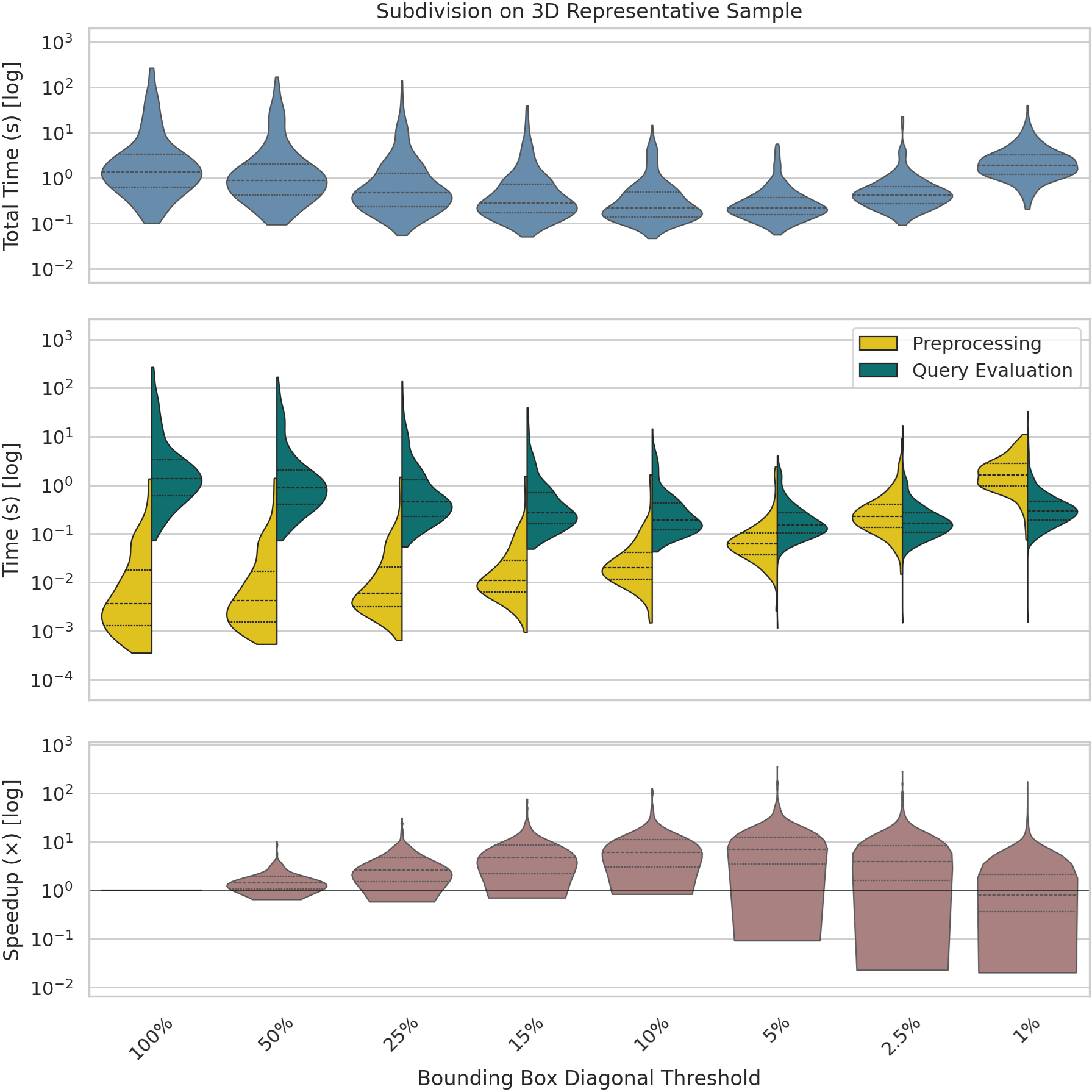} 
    \end{tabular}
    \caption{On our representative samples of 2D SVG shapes (left) and 3D CAD shapes (right), we compare the distribution of evaluation times across subdivision thresholds. 
		We show the total evaluation time (top), the evaluation time broken down by preprocessing and query evaluation (middle), 
		and the total speedup from the relevant baseline (\bezier-extracted in 2D, no subdivision in 3D). 
		The $25^\text{th}$, $50^\text{th}$, and $75^\text{th}$ percentiles are shown on each distribution. 
    All tests were performed with 14 OMP threads on a $500\times500$/$50\times50\times50$ grid of query points in an AABB of the input using a second order Taylor expansion and $\beta = 2$.}
    \Description[Timing distributions across subdivision thresholds]{Timing distributions across subdivision thresholds}
    \label{fig:subdivision_timing}
\end{figure*}

We first demonstrate the utility of our adaptive subdivision routine on each representative sample in Figure~\ref{fig:subdivision_timing}.
Each example is evaluated with 14 OMP threads on a uniform grid of $500\times500$ points in 2D and a uniform grid of $50\times50\times50$ points in 3D, with each point set generated on an AABB of the shape.
We compute the total evaluation time for each shape across various subdivision thresholds using $\beta=2$, and present the total evaluation time and a breakdown of preprocessing vs. query evaluation time, as well as speedup relative to the relevant baseline.
We note that in this experiment, each baseline still uses the proposed agglomeration algorithm, and we are comparing the processing times with different subdivision threshold.
In both cases, the equivalent \textit{direct} evaluation is often much slower.

In 2D, we see that the overall performance distribution does not significantly change with increased subdivision, despite there being virtually no overhead associated with preprocessing such cases.
This is largely because our SVG and NURBS curve parser automatically performs \bezier\ extraction, which, in the vast majority of cases, provides sufficient refinement. 
Furthermore, our implementation of~\citet{spainhour_24_robustcontainment2d} utilizes memoized subdivisions for each curve, which largely absorbs the benefits of subdividing as preprocessing. 
Nevertheless, we do see that many individual examples benefit from subdivision, some significantly.
Because subdivision also safeguards against otherwise adversarial input, such as cases with a few large elements and many more small details in a local region, we subdivide 2D input with a 10\% AABB diagonal threshold.

In 3D, however, we see that this subdivision strategy is crucial for algorithm performance. 
Our preferred AABB diagonal threshold of 10\% reduces overall runtime by an order of magnitude in many cases.
At the same time, we see that overzealous subdivision can negate this performance improvement as preprocessing time begins to dominate computation.

\begin{figure*}[t]
    \centering
    \begin{tabular}{@{}c@{}c@{}c@{}c@{}c@{}}
        \includegraphics[trim={0      0 0.1cm 0},clip,height=2.825cm]{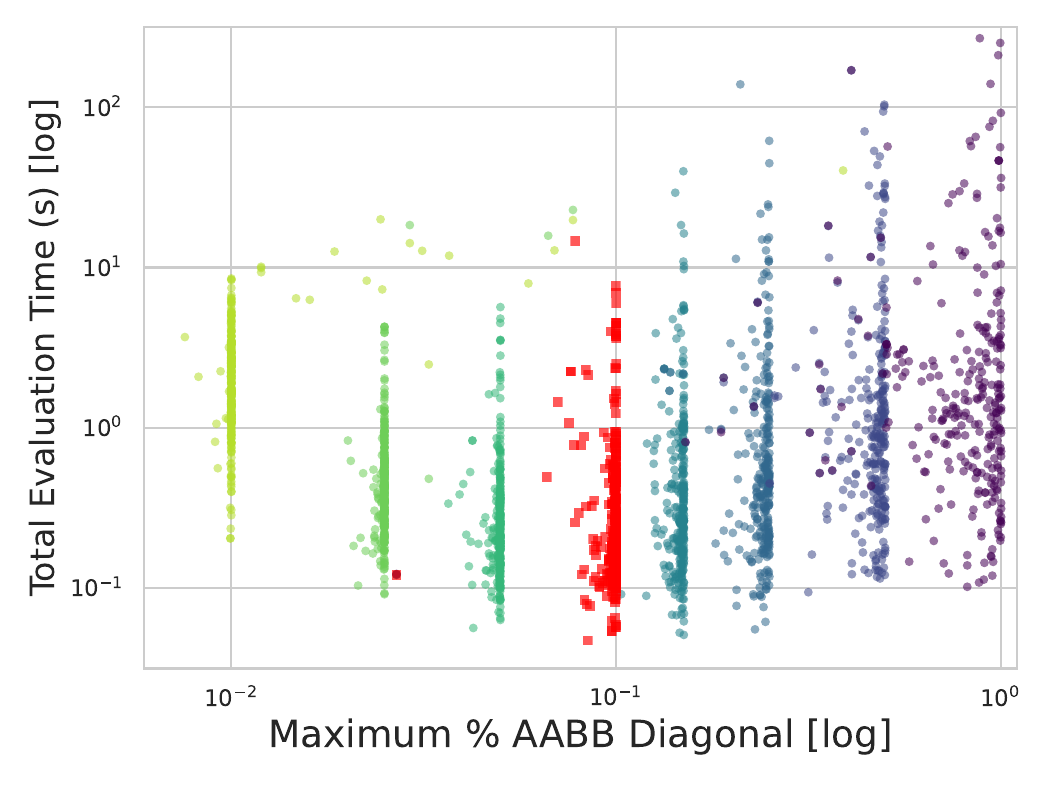}       &
        \includegraphics[trim={1.05cm 0 0.1cm 0},clip,height=2.825cm]{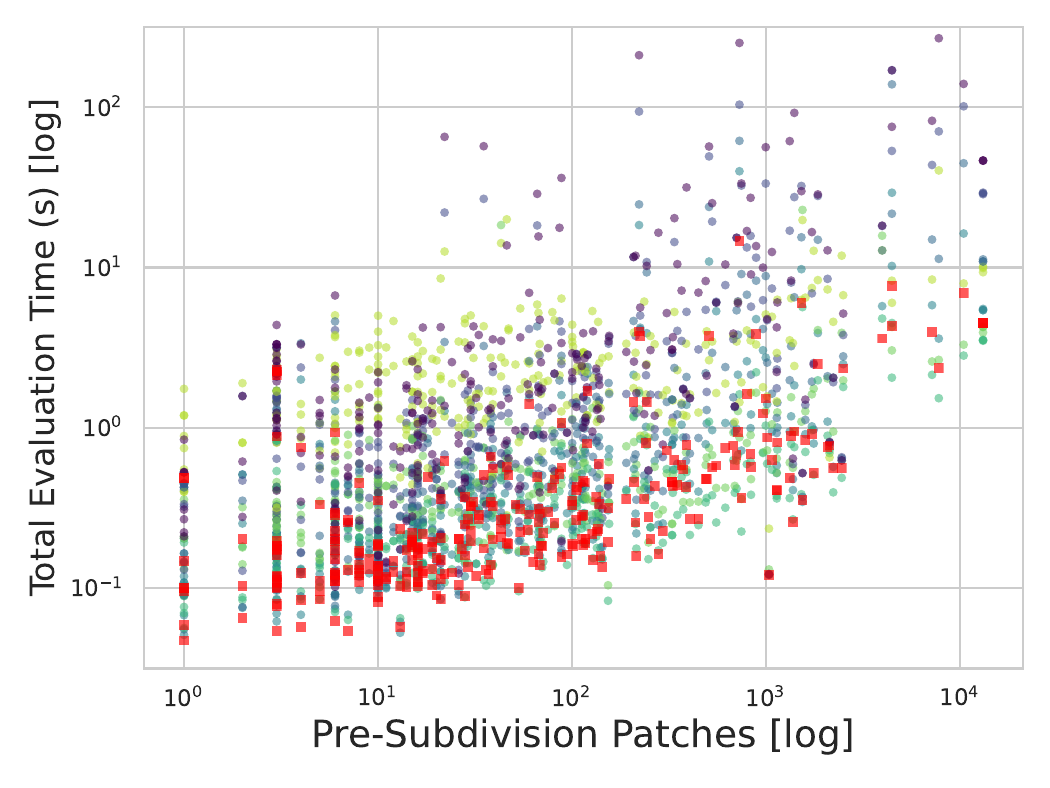}  &
        \includegraphics[trim={1.05cm 0 0.1cm 0},clip,height=2.825cm]{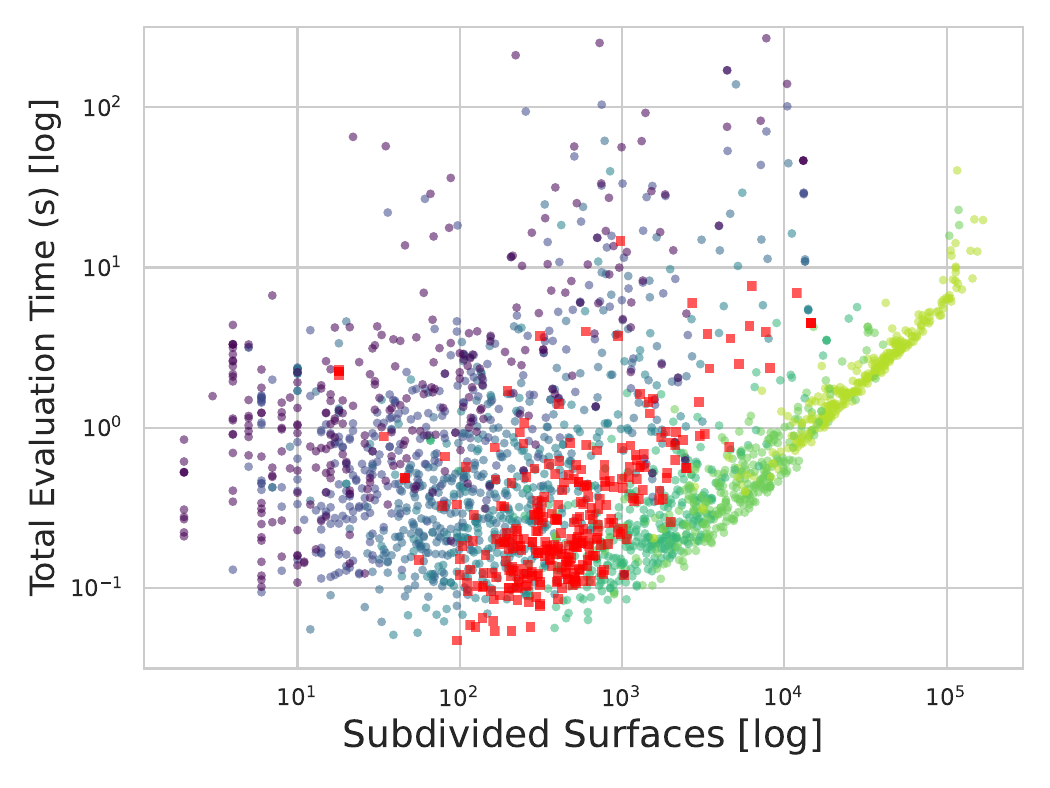}                &
        \includegraphics[trim={1.05cm 0 0.1cm 0},clip,height=2.825cm]{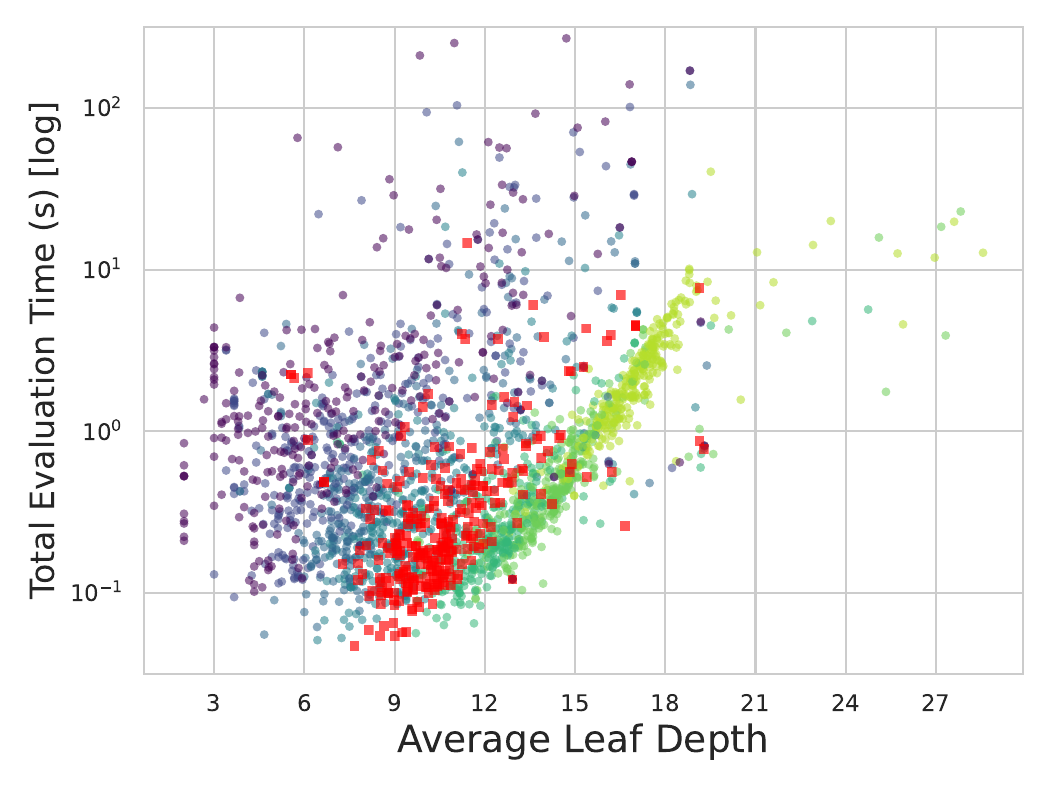}               &
        \includegraphics[trim={1.05cm 0 0     0},clip,height=2.825cm]{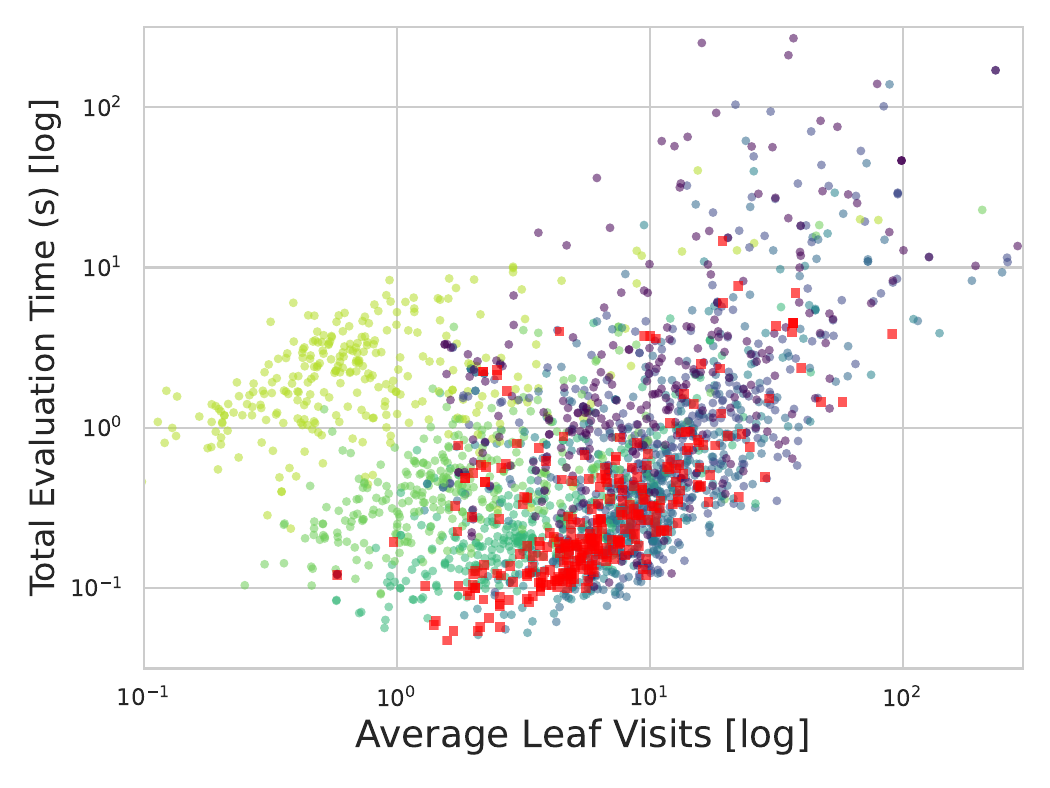}  \\
        \multicolumn{5}{c}{\includegraphics[width=0.5\linewidth]{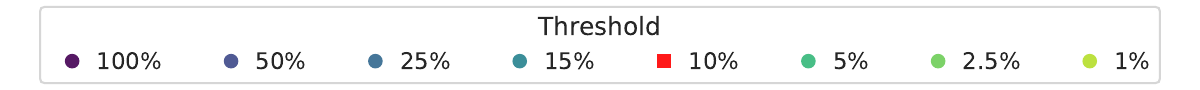}}
    \end{tabular}
    \caption{On the same representative set of 3D CAD shapes from Figure~\ref{fig:subdivision_timing}, we consider runtime performance as a function of various metrics. Each point represents one CAD shape evaluated for one subdivision threshold. The points highlighted in red represent the proposed 10\% threshold for AABB diagonal.}
    \Description[Subdivision threshold scatter plots]{Subdivision threshold scatter plots}
    \label{fig:subdivision_scatter}
\end{figure*}

To more closely consider the effect of our subdivision heuristic on 3D input, we take the  representative sample of STEP files, subdivide each according to different thresholds, and in Figure~\ref{fig:subdivision_scatter} plot algorithm performance as a function of other quantitative characteristics of both the input shape and its derived BVH.

First, we plot the timings against the maximum AABB diagonal percentage directly, which shows the same performance distribution as Figure~\ref{fig:subdivision_timing}, as expected.
By plotting timings against the number of input patches, we see no strong correlation, and only that our chosen 10\% threshold tends to minimize performance for a given shape among other thresholds.
Furthermore, even though there is a much more clear trend between the number of subdivided surfaces and performance, there is no simple ``optimal'' number of subdivisions which should be targeted during preprocessing.
Indeed, we see that our 10\% threshold generally performs as much subdivision as is advantageous for overall performance.

By plotting these timings against statistics of the resulting BVH, we similarly observe that maximizing BVH depth does not necessarily minimize performance of our method, and that our chosen threshold generally finds an appropriate balance between depth and cost. 
The same trend is observed if we instead consider the utilization of the tree via the average number of leaf nodes visited by each query. 
Even though there is eventually a point at which most queries visit \textit{no} leaf nodes, and therefore do not use the direct evaluation formula at all, enforcing this during BVH construction will not necessarily result in optimal performance.

\subsection{Algorithm Precision}\label{sec:results_precision}
We next consider the impact of agglomeration on the precision of the resulting GWN field.
We measure precision through the absolute error relative to a \emph{direct} evaluation method, for which the GWN is independently computed for each B-Rep component to the precision afforded by the analogous leaf-node evaluation method.
In 2D, this direct method has near-machine precision error~\cite{spainhour_24_robustcontainment2d}.
On the other hand, the precision of the equivalent leaf-node method in 3D is based on a user-controlled parameter, $\epsilon_q$, for an adaptive quadrature method. 
We set this parameter as $\epsilon_q = 10^{-6}$ when used in the proposed algorithm, and $\epsilon_q = 10^{-16}$ when used in the direct method, both of which are orders of magnitude below the expected approximation error.

\begin{figure}[t]
  \includegraphics[width=\linewidth]{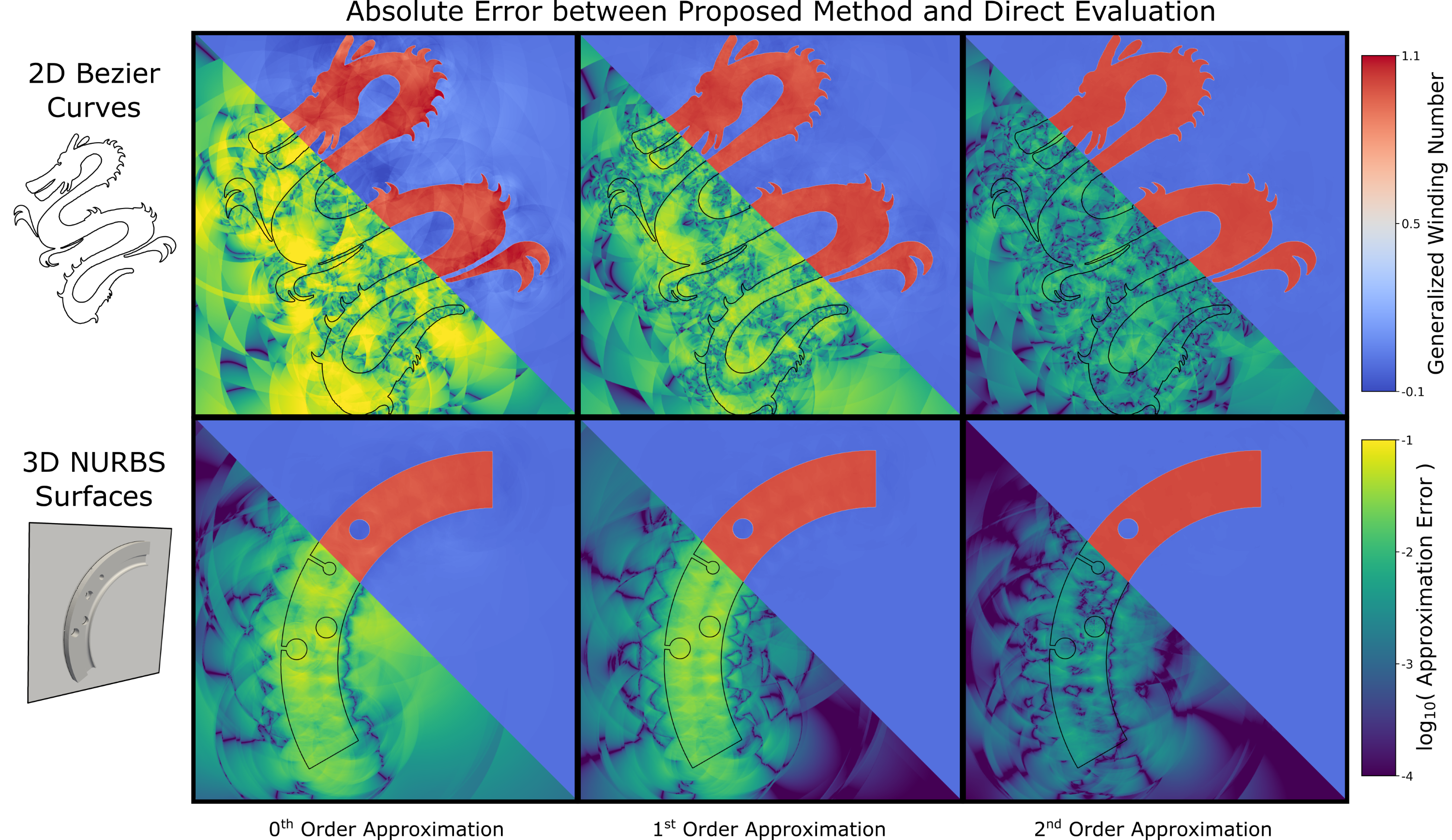}
  \caption{We show the results of our agglomeration strategy across the order of the Taylor expansion, for watertight shapes in 2D and 3D
  and the absolute error (plotted on a log scale) with respect to the direct evaluation method applied to all curves or surfaces in the shape.}
  \Description[Approximation error by Taylor order]{Taylor expansion orders for a 2D and 3D watertight example are shown. For each expansion order, the computed GWN field and the $\log_{10}$ absolute approximation error is shown.}
  \label{fig:order_comparison}
\end{figure}

As our first example, we show in Figure~\ref{fig:order_comparison} the effect of increasing Taylor expansion order on the absolute error in the GWN field for watertight shapes.
Visually, we see that increasing expansion order decreases the noise in the scalar field, and that the error decreases by roughly an order of magnitude with each expansion order.
We also see that in both examples, even the order 0 approximation keeps the error uniformly below 0.1 on the sampled points.
Importantly, this error bound indicates agreement with the direct method for a containment decision made by rounding the GWN, as disagreement on watertight shapes would require error greater than 0.5.
We explore the impact of agglomeration on containment decisions in more detail in Section~\ref{sec:results_accuracy}.

We also observe in this example that the approximation error is largest near the B-Rep itself.
Although we use direct evaluation for B-Rep components nearest to the query point, the higher concentration of \textit{other} components naturally contributes to the approximation error.
More generally, it is desirable from a performance perspective that query points can be resolved with more cluster approximations and fewer direct NURBS evaluations.
Regardless of the density of surfaces, however, the error introduced by the Taylor expansion of each cluster of B-Rep components is controlled for through the bounding sphere by which query points are considered ``far-away''.

\begin{figure}[t]
  \centering
  \begin{tabular}[t]{c}
      Parameter Sweep on 2D Sample  \\
      \includegraphics[width=\linewidth]{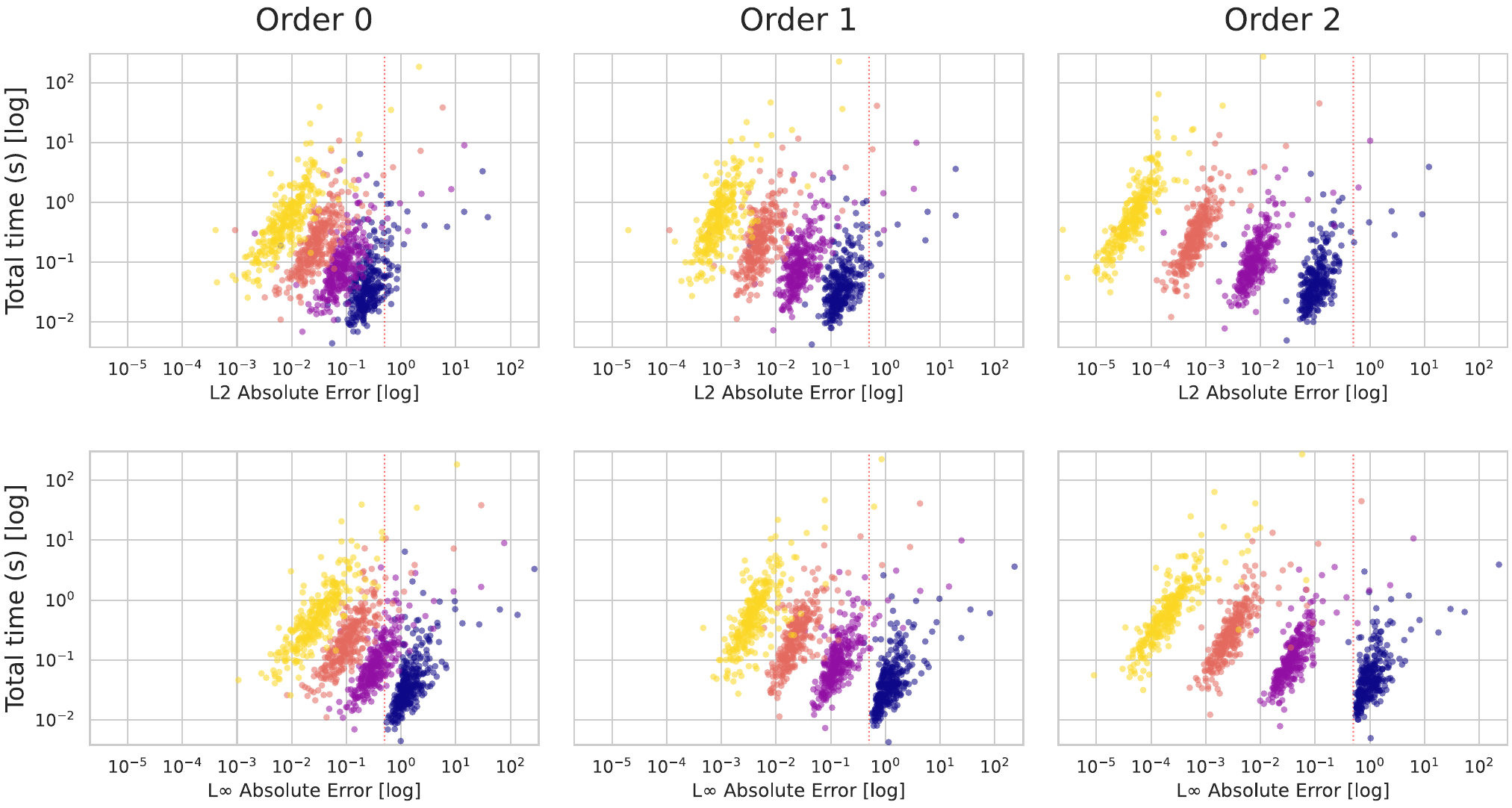} \\
      \\ Parameter Sweep on 3D Sample\\
      \includegraphics[width=\linewidth]{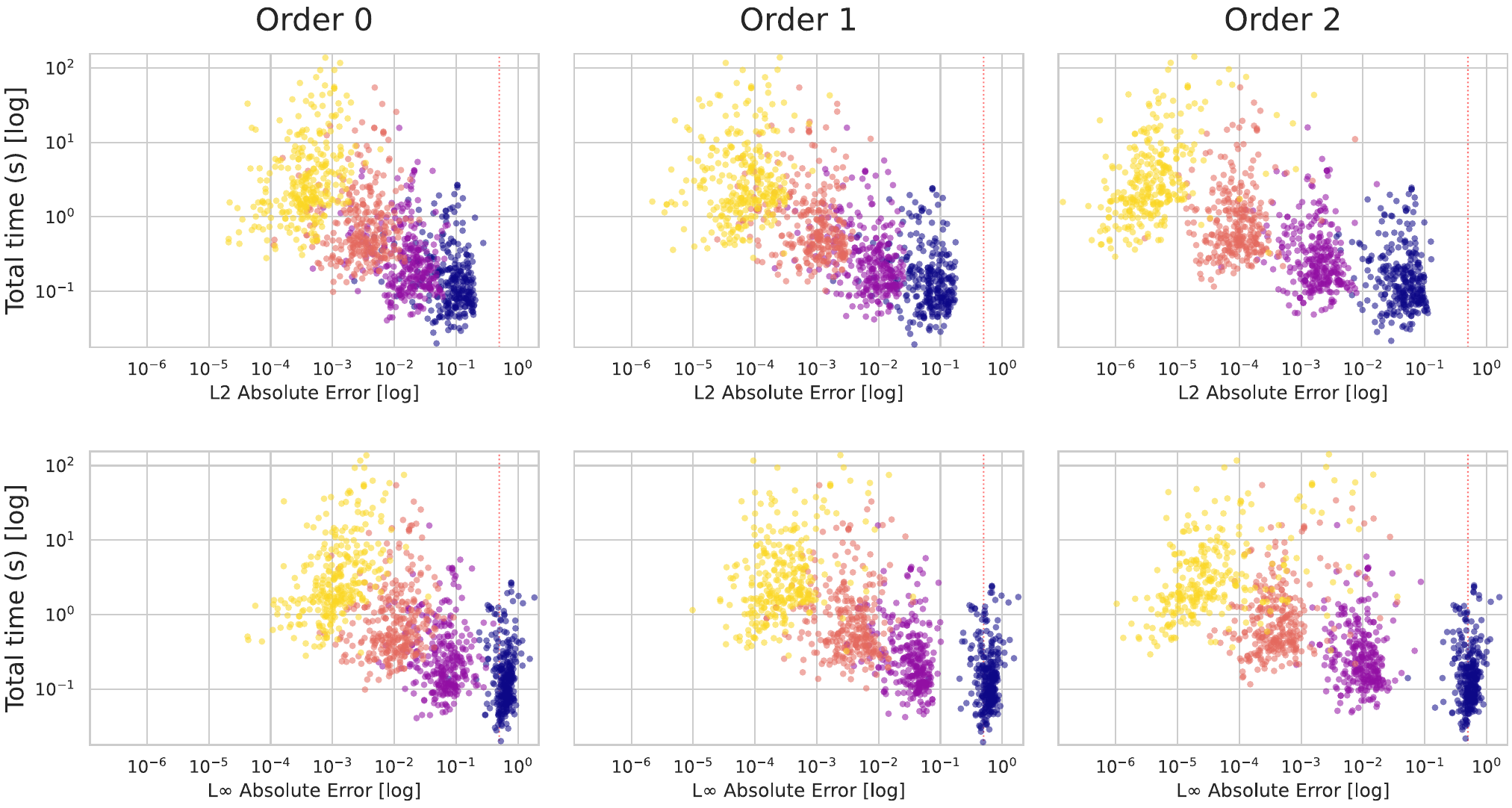} \\
    \includegraphics[width=0.4\linewidth]{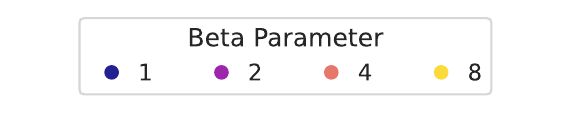}
  \end{tabular}
  \caption{On our representative samples of 2D SVG and 3D STEP files, we evaluate our agglomeration method across different Taylor expansion orders and $\beta$ parameters. For each shape, order, and $\beta$, we evaluate the GWN with 14 OMP threads on a regular grid of $1000\times1000$/$50\times50\times50$ points in 2D/3D and record the total evaluation time against the $L^2$ and $L^\infty$ absolute errors across query points.
	In each plot, errors increase along the x-axis and a red vertical line marks an error threshold of $0.5$.
		}
  \Description[Runtime versus error for different $\beta$]{Runtime versus error for different $\beta$}
		\label{fig:beta_sweep}
\end{figure}

As in~\citet{Barill-18-soupcloud}, this ``far-away'' criterion is controlled by a $\beta$ parameter, which scales the diameter of each node in the BVH (see Algorithm~\ref{alg:agglomerated_gwn}).
Increasing $\beta$ increases both the precision and runtime of the evaluation, as more points are considered near each node, and therefore fewer cluster approximations are used overall.
We see this in Figure~\ref{fig:beta_sweep}, where we evaluate the GWN on each of our representative samples for various values of $\beta$ and Taylor expansion order.
From these results, we also see that the amount by which the precision increases for larger $\beta$ itself increases with expansion order.
In other words, larger values for $\beta$ are more useful when used alongside higher-order Taylor expansions.

\begin{figure}[t]
  \centering
  \begin{subfigure}[t]{\linewidth}
    \centering
    \includegraphics[width=\linewidth]{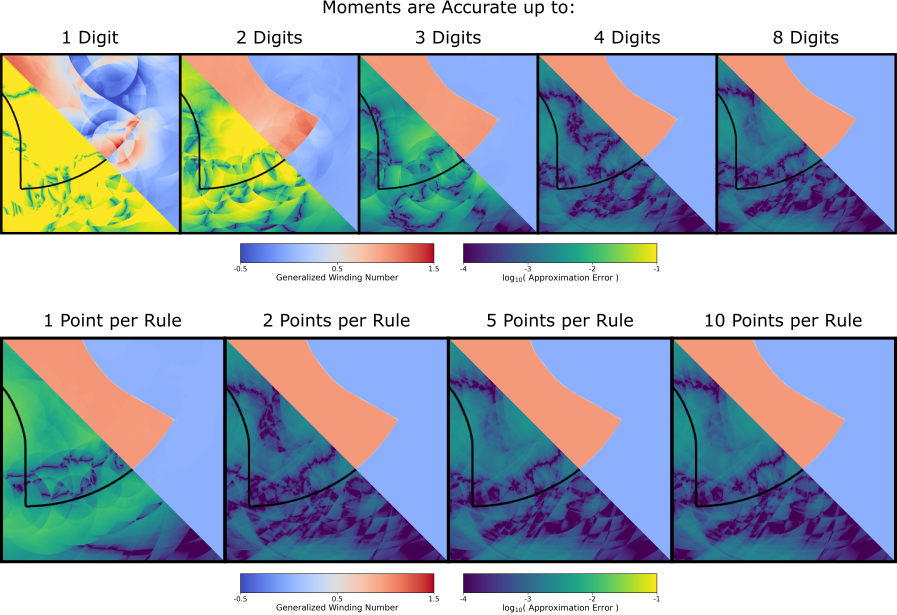}
  \end{subfigure}\\
  \begin{subfigure}[t]{\linewidth}
    \centering
    \begin{tabular}[t]{@{}l@{\hspace{0.6cm}}cc@{}}
      \toprule
      \multirow[b]{2}{*}{\makecell[l]{Points per\\Gaussian Rule}} & \multicolumn{2}{c}{Evaluation Time [\% for Moment Calculation] } \\
      \cmidrule(lr){2-3}
      & Serial (s) & Parallel (s) \\
      \midrule
      1 Point   & 2.326$\quad$ [0.005\%] & 0.295$\quad$ [0.051\%]\\
      2 Points  & 2.316$\quad$ [0.011\%] & 0.291$\quad$ [0.050\%]\\
      5 Points  & 2.314$\quad$ [0.050\%] & 0.289$\quad$ [0.059\%]\\
      10 Points & 2.336$\quad$ [0.180\%] & 0.291$\quad$ [0.144\%]\\
      \bottomrule
    \end{tabular}
  \end{subfigure}
  \caption{On a 2D slice of the 3D shape in Figure~\ref{fig:quadrature_demo}, we consider the influence of moment calculation via numerical quadrature on the overall performance and accuracy of the proposed GWN algorithm. (top) We consider the sensitivity of the Taylor approximation to moment calculation by truncating each computed moment value to varying numbers of digits, and plotting the error in the GWN field relative to the direct evaluation baseline. (bottom) We observe that even for high orders of accuracy for Gaussian quadrature, the cost of our algorithm is dominated by query time. }
  \Description[Effect of moment accuracy and quadrature order]{The top row evaluates the proposed GWN method by how many digits of the precomputed moments are kept (1, 2, 3, 4, and 8). The bottom row does the same for the number of points per Gaussian quadrature rule (1, 2, 5, and 10). For each configuration, the plotted slice is split diagonally to show the generalized winding number field in one region and the $\log_{10}$ absolute error.}
  \label{fig:quadrature_accuracy}
\end{figure}

We also observe only a slight increase in computational cost as we increase expansion order.
In general, our use of a second order expansion incurs much less computational burden than reported for triangle soups and clouds by~\citet{Barill-18-soupcloud}, 
as we typically have orders of magnitude fewer BVH nodes to track for our curved geometry.
Additionally, the cost of precomputing the necessary higher-order moments, even via numerical integration, is quickly dominated by evaluation time (see Figure~\ref{fig:quadrature_accuracy}).

Finally, this experiment also reveals important distinctions between our 2D and 3D representative samples. 
In particular, the 2D cases generally have a higher absolute error for the same value of $\beta$, despite the overall trend remaining quite consistent.
We partially attribute this to the fact that the typical 2D SVG file often is an illustration with many overlapping curves, rather than a singularly bound region.
This results in ground-truth GWN values that span a wider range than would be observed in the typical 3D STEP file in the ABC-Dataset.
In contrast, the typical 3D STEP file often has a single mechanical part that is designed to be closed, even if the result is numerically non-watertight.
In general, as the range of ground-truth GWN values widens, the linearity in the GWN and its local approximation via Taylor expansion leads to a proportional increase in absolute error, which has important consequences when the approximation is rounded to determine containment. We explore this further in Section~\ref{sec:results_accuracy}.

\begin{figure*}[ph]
  \includegraphics[width=0.825\linewidth]{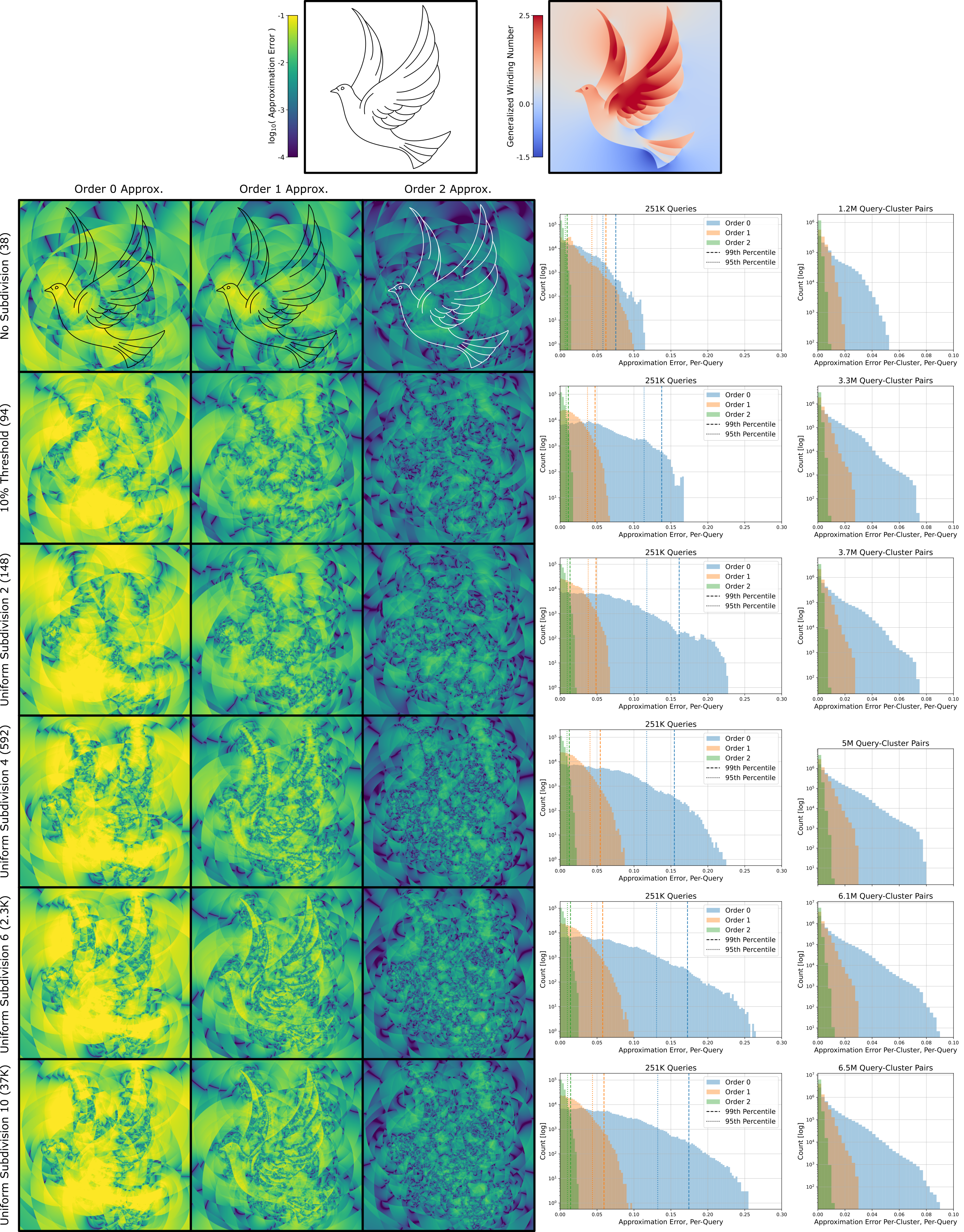}
  \caption{Using a 2D shape composed of 38 \bezier\ curves, we compare the absolute error of our agglomeration strategy with $\beta=2$ across Taylor expansion orders for different parametric refinements of the same geometry.
  We take the original shape, the shape with our 10\% maximum AABB diagonal subdivision threshold, and the original shape with varying levels of bisection applied to each curve, and evaluate the approximation on each on a grid of $500\times500$ points.
  Across these same points, we also show the distribution of errors of each internal node (for which Taylor expansion approximates the GWN of the contained curves).
  Each histogram is plotted on a logarithmic scale in frequency.
  Because this distorts the relative frequency of outliers, we also plot the 95$^\text{th}$ and 99$^\text{th}$ percentiles of the approximation error.}
  \Description[Subdivision level versus error distribution]{On the left, the absolute error in the GWN field for different subdivision settings (from no subdivision through increasingly fine uniform subdivision) and different Taylor expansion orders 0, 1, and 2 is shown. On the right, distributions for per-query approximation error and per-cluster approximation error across different expansion orders is shown.}
  \captionsetup{font=small}
  \label{fig:big_subdivision_demo}
\end{figure*}

While subdividing our parametric input as preprocessing has no effect on the ground truth GWN, it does nominally impact the precision of our proposed method.
We explore this relationship in Figure~\ref{fig:big_subdivision_demo}, where we observe that although the overall distribution of error changes as the number of curves increases, this behavior stabilizes as the domain becomes saturated with approximating clusters. 
Furthermore, we see through the associated distributions of total and per-cluster error that both remain reasonably fixed across subdivision levels, as the maximum error in each individual approximation is controlled by the choice of $\beta$.
Interestingly, we also see that for the zeroth and first order expansions, there is a stronger spatial correlation between the B-Rep and the error. 
Because the result is more robust to subdivision, paired with the marginal increase in cost, we use this second order expansion in the presented numerical results.

Finally, we consider how our use of numerical quadrature in moment calculation for 3D shapes impacts the overall precision of the proposed algorithm. 
We observe in Figure~\ref{fig:quadrature_accuracy} that while moments must be reliably computed to be useful, 
our method is largely insensitive to the number of quadrature points used in the method of~\citet{gunderman-21-trimmednurbsintegration}, both in terms of performance and precision. 
This can largely be attributed to the many levels of subdivision applied to the original input shape, namely surface subdivision by maximum AABB diagonal as preprocessing, and \bezier\ extraction of NURBS surfaces and trimming curves during moment evaluation. 
Furthermore, we observe in the associated table that even with highly precise quadrature rules, the cost of our algorithm is dominated by query time.

\subsection{Algorithm Accuracy}\label{sec:results_accuracy}
An important goal for our fast GWN method for curved geometric objects is an accurate containment query at arbitrary locations in the domain.
This is largely ensured by our choices for direct evaluation method, which uses the exact B-Rep representation to correctly place query points on the appropriate side of the boundary.
Since we determine containment by rounding the GWN, our agglomeration algorithm requires careful control of the overall approximation in the GWN field.

\begin{figure}[h]
  \includegraphics[width=\linewidth]{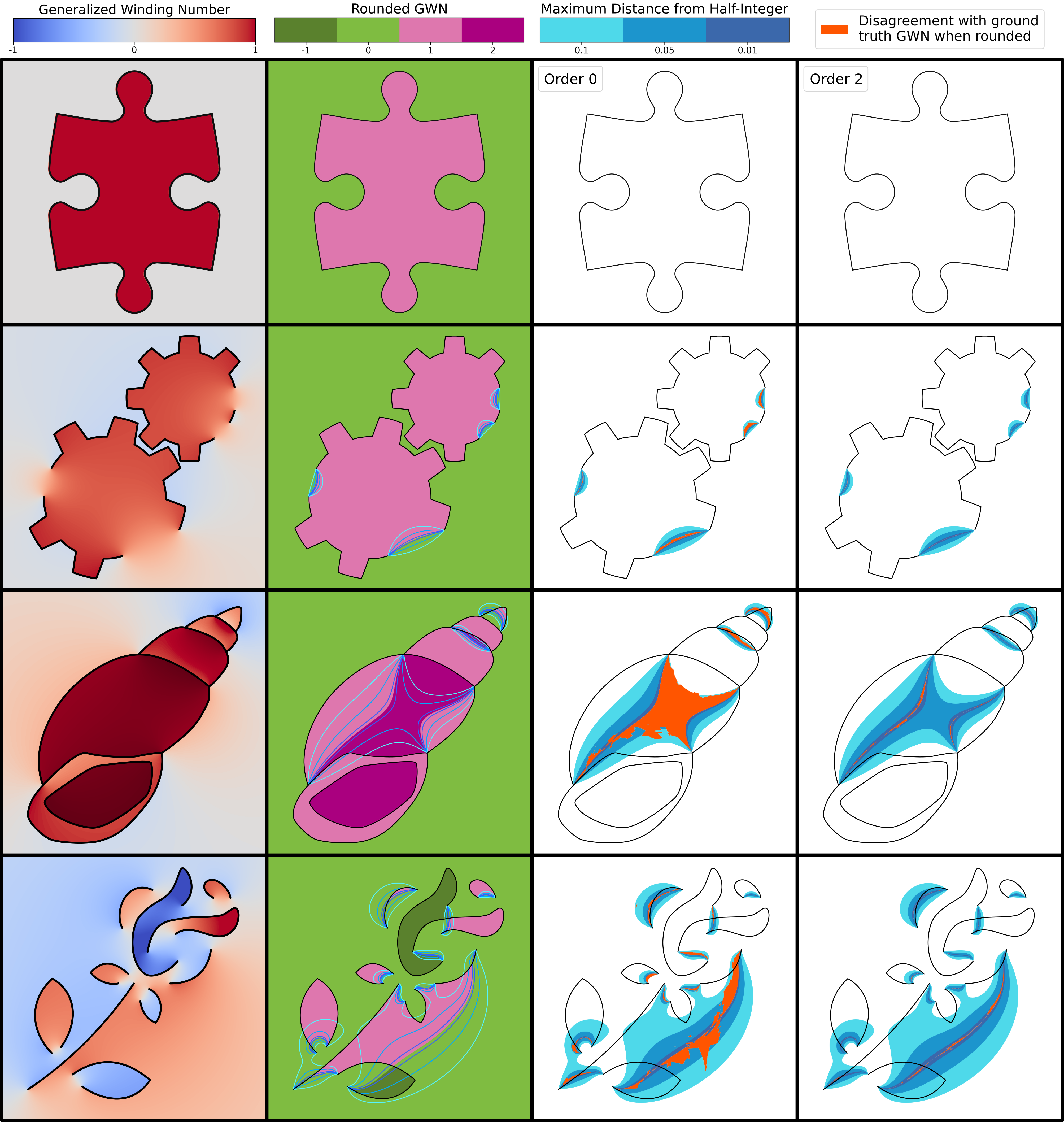}
  \caption{We consider the utility of the rounded GWN scalar field as a proxy for containment on shapes which are increasingly non-watertight, and how the proposed agglomeration technique impacts the rounded GWN field. (left) We show the ground-truth and rounded GWN, alongside contours which mark regions for which the ground-truth GWN is within a certain value of a half-integer. (right) For an order 0 and order 2 Taylor expansion, we mark all points for which there is disagreement between the ground-truth GWN and the approximated GWN, superimposed over proximity of the ground-truth GWN to a half-integer.}
  \Description[Containment via rounded GWN]{A 4-by-4 grid of examples compares ground-truth winding numbers, rounded containment, and regions where the rounded approximation disagrees with the ground truth. The first column shows the continuous GWN field with the shape outline overlaid. The second column shows the rounded winding number, along with contours that show distances to the nearest half-integer. The third and fourth columns show, for order 0 and order 2 approximations respectively, overlays marking pixels where rounding the approximation disagrees with rounding the ground truth.}
  \label{fig:disagreement_demo}
\end{figure}

Naturally, any amount of precision error can be sufficient to cause the rounded approximate value to differ from the rounded ground-truth GWN. 
Indeed, if the ground-truth GWN is exactly $0.5$, then any amount of approximation error will result in \textit{disagreement} between the two rounded values. 
However, such disagreement does not necessarily reflect a genuine \textit{misclassification} with respect to the bounding geometry of the shape.
A common interpretation of the GWN is a measure of confidence that the containment decision made through rounding is reasonable~\cite{Jacobson-13-winding}, in the sense that a point for which the GWN is 0.9 is more meaningfully ``inside'' the provided geometry than one for which the GWN is 0.6, even if both round to the same value.
From this perspective, the true containment values for a shape are increasingly ambiguous as the GWN approaches half-integers, and so there is no singularly correct containment decision which can be made in the first place.
Furthermore, as the shape becomes less watertight and more abstract (a case particularly common for the SVG shapes in OpenClipArt20k), the region of space with ambiguous containment grows.
We depict such cases in the first two columns of Figure~\ref{fig:disagreement_demo}.

Within this framework, we claim that our proposed method, under an appropriate parameterization, reliably returns an accurate containment decision whenever there is an unambiguous decision to make. 
This means that while we do not guarantee that containment decisions around half-integers will always match that of the rounded ground truth, we maintain a level of approximation error that permits a rounding-based containment query at all other points.

Moreover, we always respect the jump condition in the GWN field that exists around the provided boundary elements.
We see this in the third and fourth columns of Figure~\ref{fig:disagreement_demo}, where all disagreements occur exactly near where the GWN field takes half-integer values rather than across the boundary curves themselves.
This figure also demonstrates that as the order of the Taylor expansion increases, the total amount of disagreement decreases commensurately, and their distribution narrows around the half-integer contour.

In a containment context, any absolute error greater than 0.5 will guarantee a misclassification.
Based on Figure~\ref{fig:beta_sweep}, we set $\beta = 2$ for our 3D examples from the ABC-Dataset, and $\beta = 4$ for our 2D examples from the OpenClipArt20k dataset. 
As described in that section, we have found this more conservative value for $\beta$ to be necessary for ensuring reliable containment decisions on these 2D SVG examples despite the marginally increased cost, as we must compensate for the increased frequency of examples within that dataset that have large amounts of overlapping or nearly-overlapping curves.
Such overlaps do not meaningfully influence the qualitative behavior of the relevant Taylor expansion, but do scale the overall range of GWN values and therefore proportionally increase the absolute error, which we can straightforwardly control with a larger $\beta$.

\begin{figure}[t]
  \centering
  \includegraphics[width=1.0\linewidth]{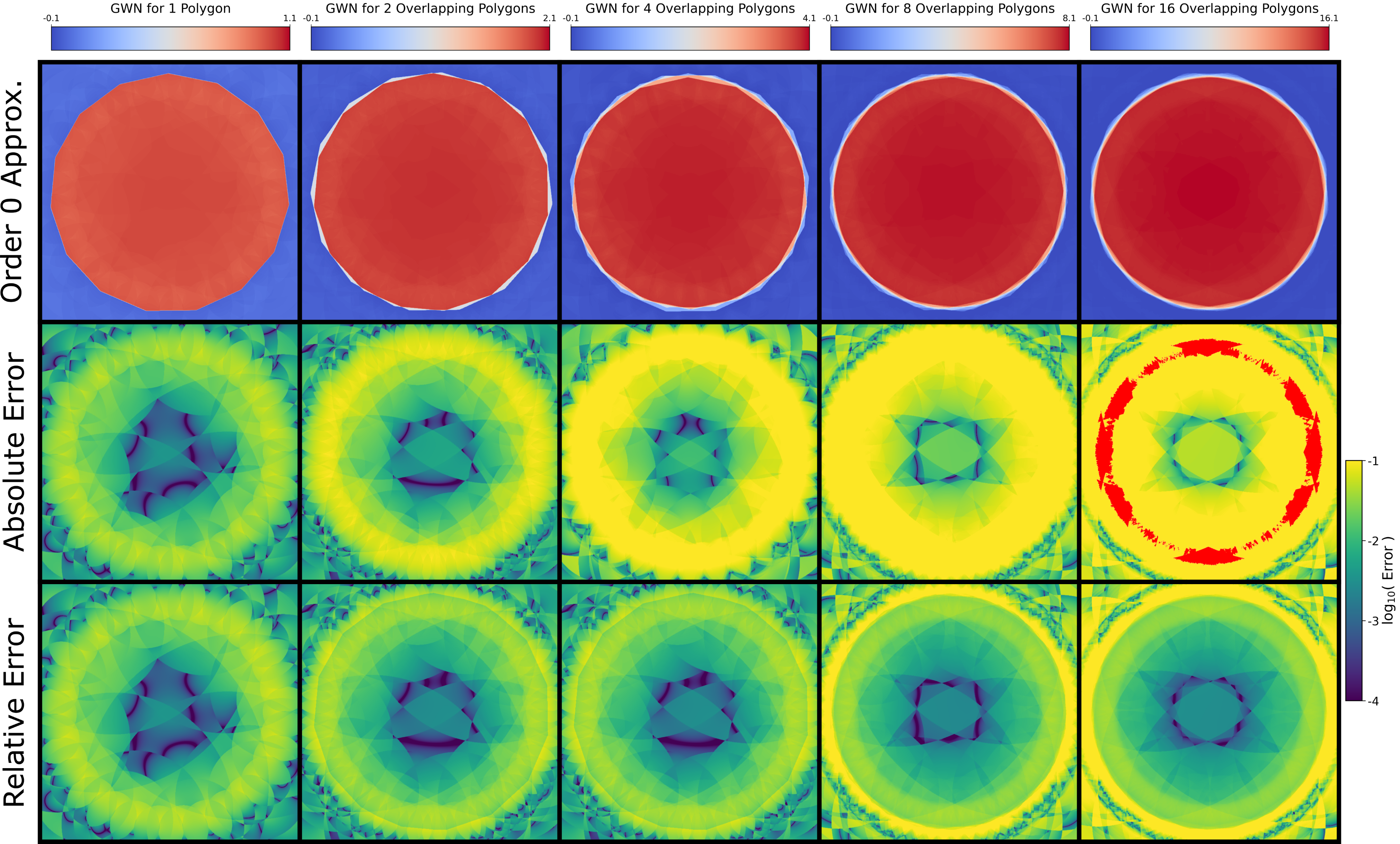}
  \caption{We evaluate the approximate GWN field computed via our agglomerated algorithm with expansion order 0 and $\beta=2$ for 1, 2, 4, 8, and 16 closed and nearly-overlapping 15-sided (linear) polygons, alongside the absolute and relative error in the approximation.
  Over the absolute error, we highlight in red each point for which rounding the approximated GWN results in a misclassification.}
  \Description[Overlapping polygons and scaling error]{The GWN for overlapping polygons (1, 2, 4, 8, and 16) is shown. The top row shows the order-0 approximated generalized winding number field for each case. The middle row shows $\log_{10}$ absolute error, with misclassifications marked in red. The bottom row shows $\log_{10}$ relative errors for the same five cases.}
  \label{fig:overlap_demo}
\end{figure}

We isolate this effect in Figure~\ref{fig:overlap_demo}, where we overlap increasingly many 15-sided, closed polygons and evaluate the GWN field for each using $\beta=2$ and the $0^{\text{th}}$ order Taylor expansion (the lowest order available). 
Because the input is watertight, the ground-truth GWN is integer valued, and the approximation from the Taylor expansion is near-integer valued.
However, each additional polygon added to the shape increases the GWN of their overlap, and while the relative error and general appearance of the GWN field remains largely unchanged, the absolute error increases linearly. 
Indeed, there is a point at which misclassifications are recorded even for this watertight shape.
By using linear edges in these overlapping polygons, we highlight that being susceptible to such error alignment is not specific to our use of curved geometry in the approximation, but rather a consequence of using Taylor expansion to approximate the GWN in the first place.

\begin{figure*}[h]
    \begin{tabular}[t]{c}
      \includegraphics[width=0.98\linewidth]{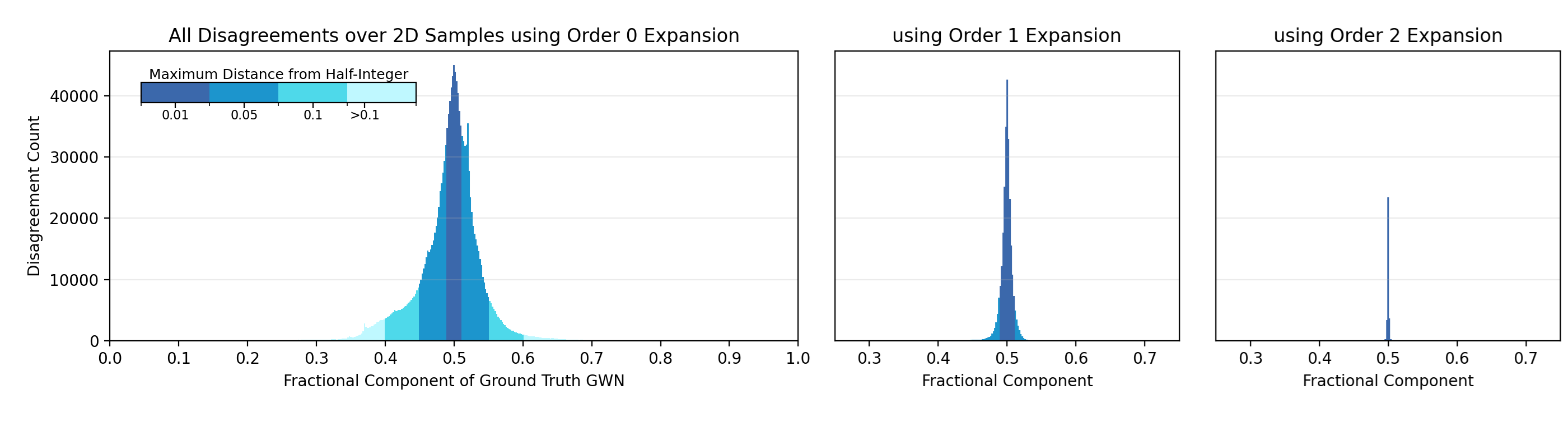} \\
      \includegraphics[width=0.98\linewidth]{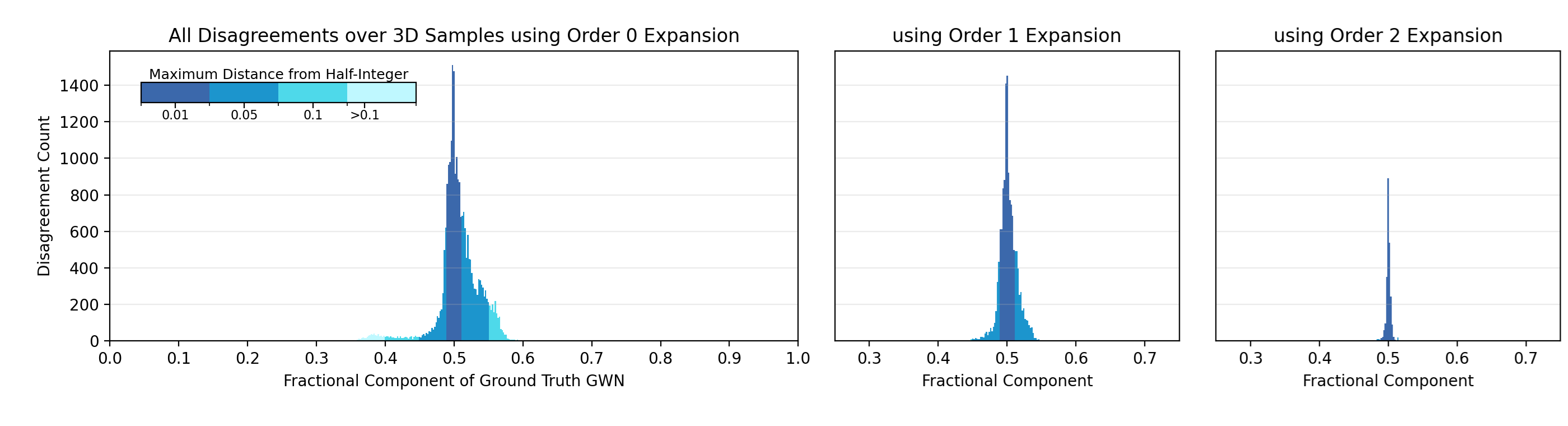}
  \end{tabular}
  \caption{Across our representative 2D and 3D datasets, we evaluate the GWN on a grid of $1000\times1000$/$50\times50\times50$ points, and record all points where the rounded approximation disagrees with the rounded ground-truth GWN. For each of these disagreements, we record the fractional component of the ground-truth GWN, and plot the distribution of disagreements as a function of these fractional components for each expansion order.}
  \Description[Disagreement histograms by fractional part]{Two stacked plots summarize where rounded disagreements occur for 2D (top) and 3D (bottom) datasets. Each plot contains three panels, one per expansion order (0, 1, and 2), showing a histogram of the fractional component of the ground-truth winding number at disagreement points. Each distribution is sharply peaked around a fractional value of 0.5.}
  \label{fig:disagreement_distribution}
\end{figure*}

Nevertheless, by selecting the appropriate $\beta$, we achieve remarkably consistent results across both datasets, even with the lowest expansion order.
We demonstrate this in Figure~\ref{fig:disagreement_distribution} by evaluating the GWN on a uniform query grid ($1000\times1000$ in 2D, $50\times50\times50$ in 3D).
We record the total number of containment disagreements among all query points on all examples, and sort them according to the fractional component of the ground-truth GWN at that point.
From the resulting distributions, we see that disagreements occur only in the vicinity of half-integers, and that the overall quantity decreases with the order of the Taylor expansion. 
Furthermore, we also see empirically that many fewer disagreements are observed on the 3D examples overall, largely reflecting the difference in typical use-case between the two categories of examples.

For this experiment, we have excluded six outliers from our representative SVG samples for which an exceptional number of nested components $(>16)$ requires an even more conservative $\beta$ to process accurately. 
A strategy which dynamically selects $\beta$ based on geometric properties of the components within a cluster represents an interesting area for potential future improvement.

\begin{figure*}[h]
  \includegraphics[width=0.85\linewidth]{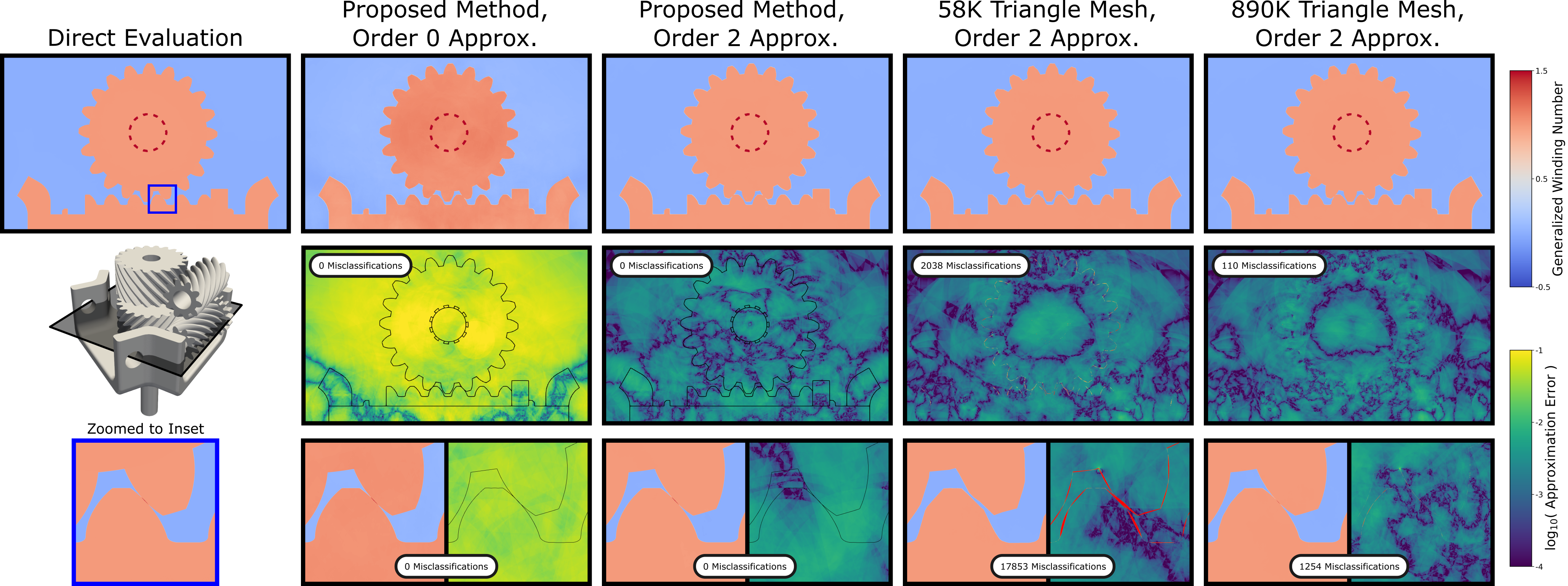}
  
  \vspace{0.5cm}

  \begin{tabular*}{\textwidth}{@{\extracolsep{\fill}} lrrrrr @{}}
    \toprule
    Primitive Type
    & NURBS Mesh
    & NURBS Mesh
    & NURBS Mesh
    & 58K Triangle Mesh
    & 890K Triangle Mesh \\
    Evaluation Method
    & Direct
    & Order 0 Approx.
    & Order 2 Approx.
    & Order 2 Approx.
    & Order 2 Approx. \\
    \midrule
    Preprocessing Time (s)   & 0.145 & 0.264 & 0.280 & 0.430 & 5.95 \\
    Query Evaluation Time (s) & 287   & 2.09  & 2.19  & 0.267 & 0.518 \\
    \midrule
    Total Time (s)           & 287   & 2.35  & 2.47  & 0.697 & 6.47 \\
    \bottomrule
  \end{tabular*}
  \captionsetup{font=small}
  \caption{We compare our proposed algorithm to triangulations of a 3D shape, whose GWN is computed using the agglomeration method of~\citet{Barill-18-soupcloud}.
  We evaluate each on uniform grids of $1000\times1000$ points on two planar slices of the shape, the first of which is clipped for brevity.
  We plot the absolute error relative to direct evaluation, and highlight in red any point with an absolute error greater than $0.5$, which indicates that a containment decision made by rounding produced an incorrect result.
  The total number of these misclassifications for each method is presented for each view.
  Below, we also record the runtime needed to evaluate the GWN on a uniform grid of $50\times50\times50$ points with 14 OMP threads, broken down by query time and preprocessing time, which in the case of triangle meshes, includes the time required for OpenCascade to compute the triangulation from the input STEP file.}
  \Description[Curved input versus triangulation]{The figure compares GWN methods for a mechanical part on planar slices: direct evaluation on the original curved model, the proposed method with order-0 and order-2 approximations, and order-2 agglomeration on two triangle meshes (approximately 58K and 890K triangles). The top row shows the generalized winding number field for each method on the slice (blue-to-red). Additional panels show the absolute error relative to direct evaluation, with red overlays indicating locations with misclassifications. A final row shows a zoomed inset region with the corresponding field and error views.}
  \label{fig:bulk_accuracy}
\end{figure*}

In contrast to the proposed method, fast GWN strategies which discretize curved input often induce unexpected, and in our case wholly undesirable containment errors.
While agglomeration of a triangulation will almost always incur less computational cost than the equivalent agglomeration of curved input, this comes with a tradeoff of degraded accuracy and the increasingly burdensome cost of generating fine triangulations of NURBS input.
For example, we see in Figure~\ref{fig:bulk_accuracy} that even at high resolutions, triangulation of a 3D CAD surface almost necessarily introduces misclassifications near the surface.
In contrast, our proposed method faithfully reproduces the curved boundary even with the lower order Taylor expansion.

\begin{figure*}[h]
  \includegraphics[width=0.85\linewidth]{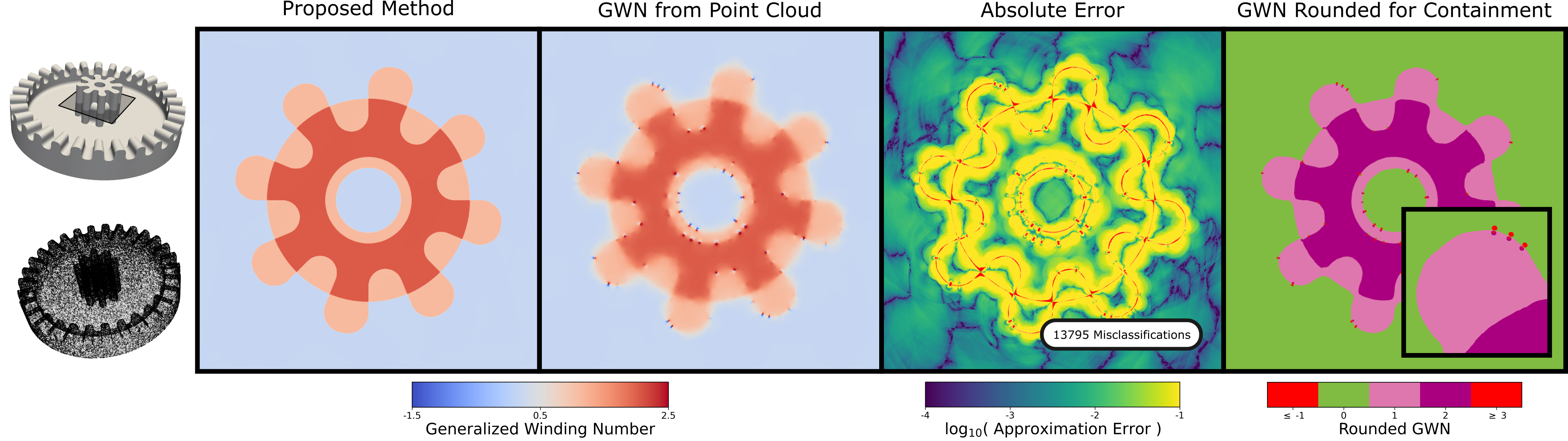}
  \captionsetup{font=small}
  \caption{We compare our proposed algorithm to the GWN of a cloud of $10^6$ points sampled uniformly from the surface of the shape, evaluated on a planar slice at $1000\times1000$ query points.
  We plot the absolute error relative to direct evaluation, highlight in red any points for which the rounded GWN differs from the rounded GWN of direct evaluation, and count the resulting number of misclassifications. 
  We also plot these rounded GWN values directly.}
  \Description[Point-cloud GWN comparison]{A rendered 3D CAD part and its sampled point cloud are shown at left, along with the planar slice used for evaluation. Shown are the proposed method's winding number field, a winding number field computed from the point cloud, the absolute error relative to direct evaluation with red markers indicating misclassified points, and a map of the rounded winding number used for containment with a zoomed inset.}
  \label{fig:cloud_gwn}
\end{figure*}

Similarly, although GWN methods for point clouds are proficient at producing reasonable isosurfaces for messy input~\cite{Barill-18-soupcloud}, we find such methods to be largely unsuitable for point clouds sampled directly from CAD input (as explored by~\citet{balu-23-immersogeometric}),
as the use of dipoles as the fundamental geometric representation of the object necessarily introduces meaningful containment disagreement.
We consider this in Figure~\ref{fig:cloud_gwn} for a shape with overlapping components, where it is impossible to identify by rounded GWN alone if some points are truly located within both pieces, or simply have a GWN perturbed by a nearby dipole.

\subsection{Algorithm Performance}

We performed large scale performance studies to properly characterize the runtime complexity of our proposed methods, primarily our fast agglomerated methods in 2D and 3D, using a second order expansion with $\beta=2$ in each case. Each method (proposed and direct; 2D and 3D) was run threaded using 14 OMP threads, corresponding to a single tile on our machine nodes.

In 2D, we compare our algorithm against the entire ${\sim}20$K SVG images in the OpenClipArt20k dataset~\cite{openclipart} using a $1000^2$ grid of sample points 
taken uniformly from the SVG \texttt{viewBox}.
For this experiment, we converted all curves to cubic \bezier\ components (rational cubic for elliptical arcs) during parsing.

In 3D, we compare our algorithm against one 10k chunk of the ABC-Dataset~\cite{koch-19-abcdataset}, which contains a total of one million in-the-wild CAD surfaces of varying quality used to benchmark algorithms related to CAD processing and machine learning models.
Within this 10k chunk of STEP files, we have discarded roughly 800 which our OpenCascade reader~\cite{opencascade} was unable to successfully parse. 
In most cases, this was due to interoperability issues with the Onshape modeling engine from which the ABC-Dataset is collected.
We evaluate performance for each model on a $50^3$ grid of sample points on a bounding box of each shape.

\begin{figure*}[tp]
  \centering

  \begin{subfigure}[t]{\textwidth}
    \includegraphics[width=.49\textwidth]{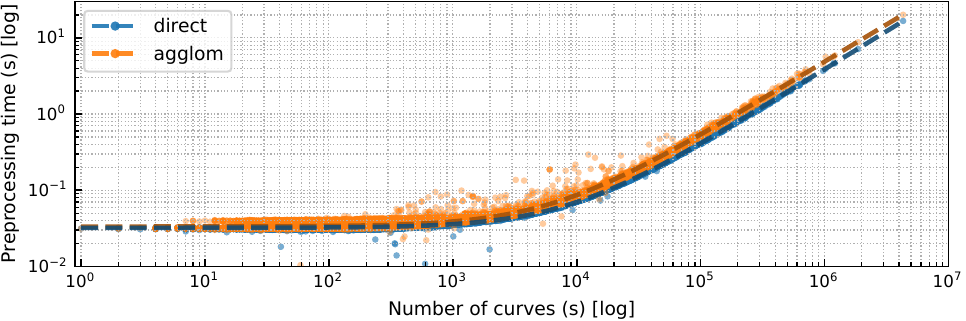}\hfill
    \includegraphics[width=.49\textwidth]{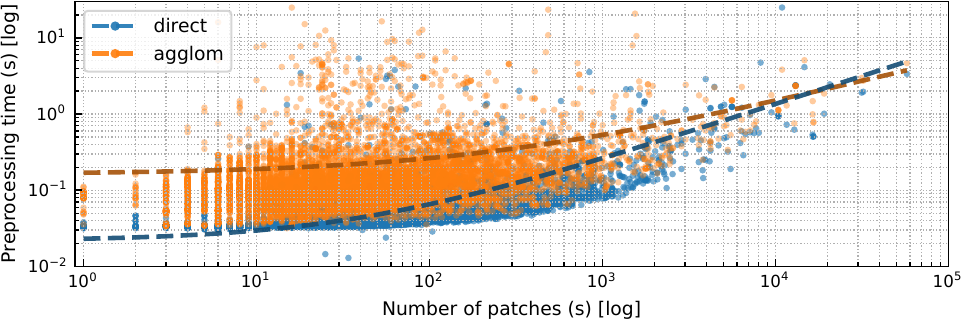}
		\caption{Preprocessing times for the 2D OpenClipArt20k dataset (left) and 3D ABC-Dataset (right) on a log-log scale.}\captionsetup{font=small}
		\label{fig:performance_preprocess}
  \end{subfigure}\\
  \begin{subfigure}[t]{\textwidth}
    \includegraphics[width=.49\textwidth]{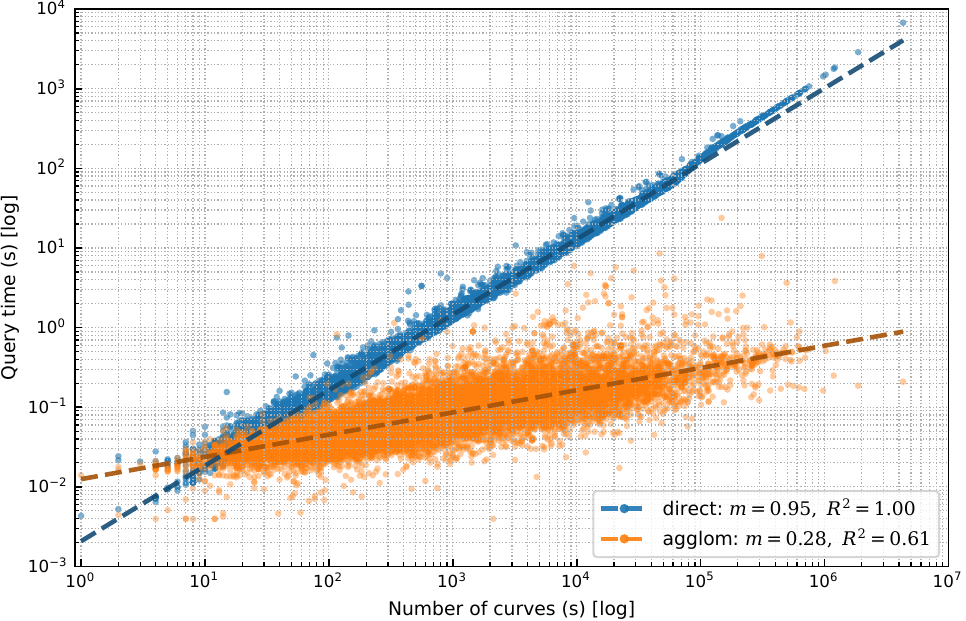}\hfill
    \includegraphics[width=.49\textwidth]{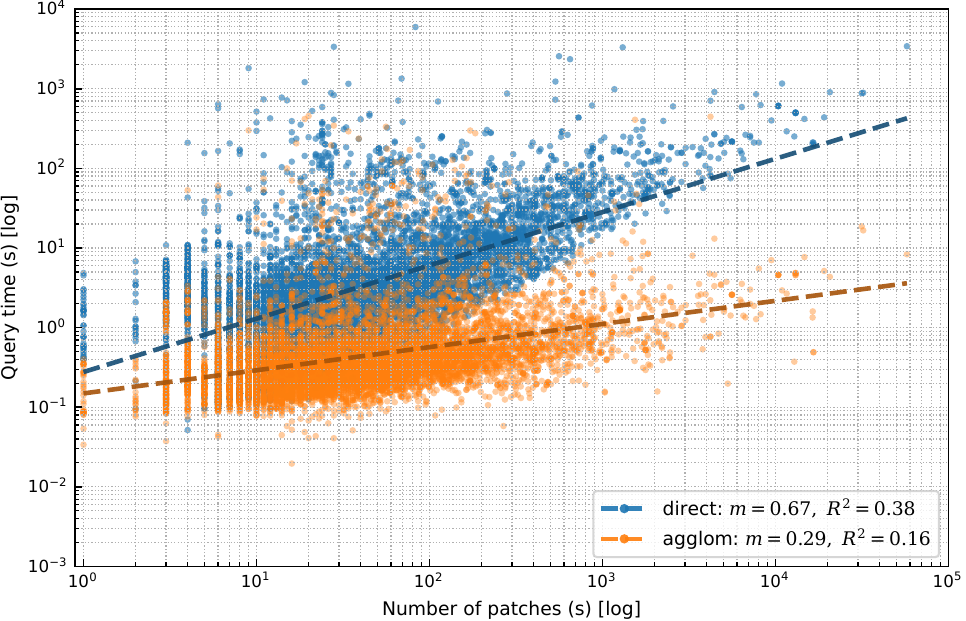}
		\caption{Query times for the 2D OpenClipArt20k dataset (left) and 3D ABC-Dataset (right) on a log-log scale. 
		         We provide the slope $m$ and the coefficient of determination $R^2$ for each best-fit line }\captionsetup{font=small}
		\label{fig:performance_query}
  \end{subfigure}\\
  \begin{subfigure}[t]{\textwidth}
    \includegraphics[width=.49\textwidth]{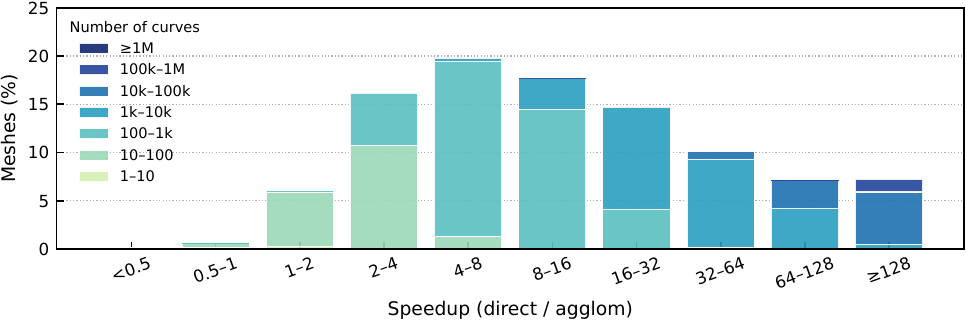}\hfill
    \includegraphics[width=.49\textwidth]{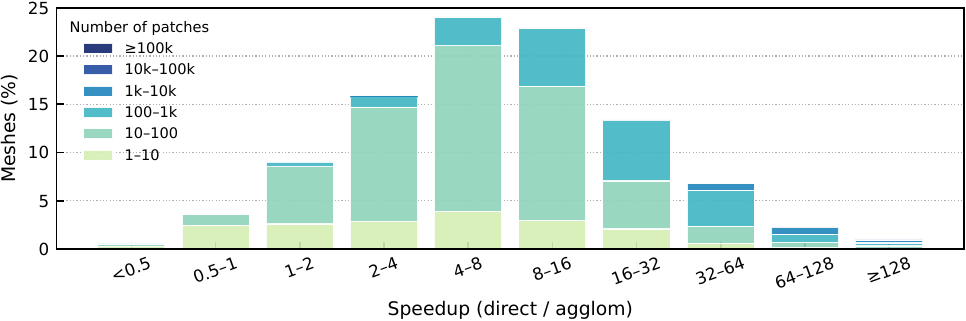}
		\caption{Stacked histogram of query speedups when using our \texttt{agglomerated} algorithm vs. \texttt{direct} across the 2D and 3D datasets.
			The histogram bars show the percent of the datasets in each specified speedup range, broken down by the number of input primitives.
		}\captionsetup{font=small}
		\label{fig:performance_speedups}
  \end{subfigure}
  \caption{Performance of our proposed algorithm in 2D (left) and 3D (right) using Taylor expansions of order 2 and $\beta=2$.
			In 2D, we compare our agglomerated algorithm (orange) against the direct algorithm (blue) on the OpenClipArt20k dataset comprised of 20,000 
			SVG images at resolution $1000^2$.
			In 3D, our comparisons were on 9,198 CAD models (one chunk of the ABC-Dataset) at resolution $50^3$.
			All experiments were run with 14 OpenMP threads.
			}
  \Description[Large scale performance plots]{Large scale performance plots}
  \label{fig:performance_experiments}

\end{figure*}

We present our overall performance improvements for the proposed agglomeration methods in Figure~\ref{fig:performance_experiments}.
In 2D, we see very strong linear scaling for the direct method, and sub-linear scaling for our proposed agglomeration strategy as expected (Figure~\ref{fig:performance_query} left).
We also see that the preprocessing necessary to use 2D input in our agglomeration strategy is virtually constant across the number of input curves, as subdivision can be evaluated rapidly for a given curve direct, and formulae are used to compute moments.
As seen in the distribution of speedup factors in Figure~\ref{fig:performance_speedups}(left), we see greater than $10\times$ improvement in more than half of the examples, and more than a $30\times$ speedup in the upper quartile.
Naturally, the speedup increases with the number of input curves.

We see similar gains in 3D, where runtime also grows sub-linearly with the number of input patches (Figure~\ref{fig:performance_query} right).
We do not expect fully logarithmic scaling in 3D, as the performance of the hierarchical approach is heavily dependent on the number of patches in the subdivision output, which is not strictly correlated to the number of patches in the original input.
We also see impressive speedups in 3D over direct evaluation. Several models which would take on the order of $10^3$ seconds with direct evaluation, run in under one second with our agglomeration method.
Preprocessing takes longer in 3D than in 2D, but remains a small fraction of the total runtime, requiring less than one second in most cases.

In both 2D and 3D, a small fraction of examples slow down with agglomeration (<1\% in 2D and <5\% in 3D),
primarily on inputs with very few input primitives (i.e., less than 10).
In 3D, this slowdown is often caused by compatibility issues between the input STEP file and our OpenCascade reader, involving very small knot spans, or trimming curves outside the bounding box of their patch, leading to overly refined preprocessing or expensive evaluation.
We plan to investigate these cases and improving our STEP reader to better support these relatively rare cases.

\begin{figure}[t]
    \centering
    \includegraphics[width=0.9\linewidth]{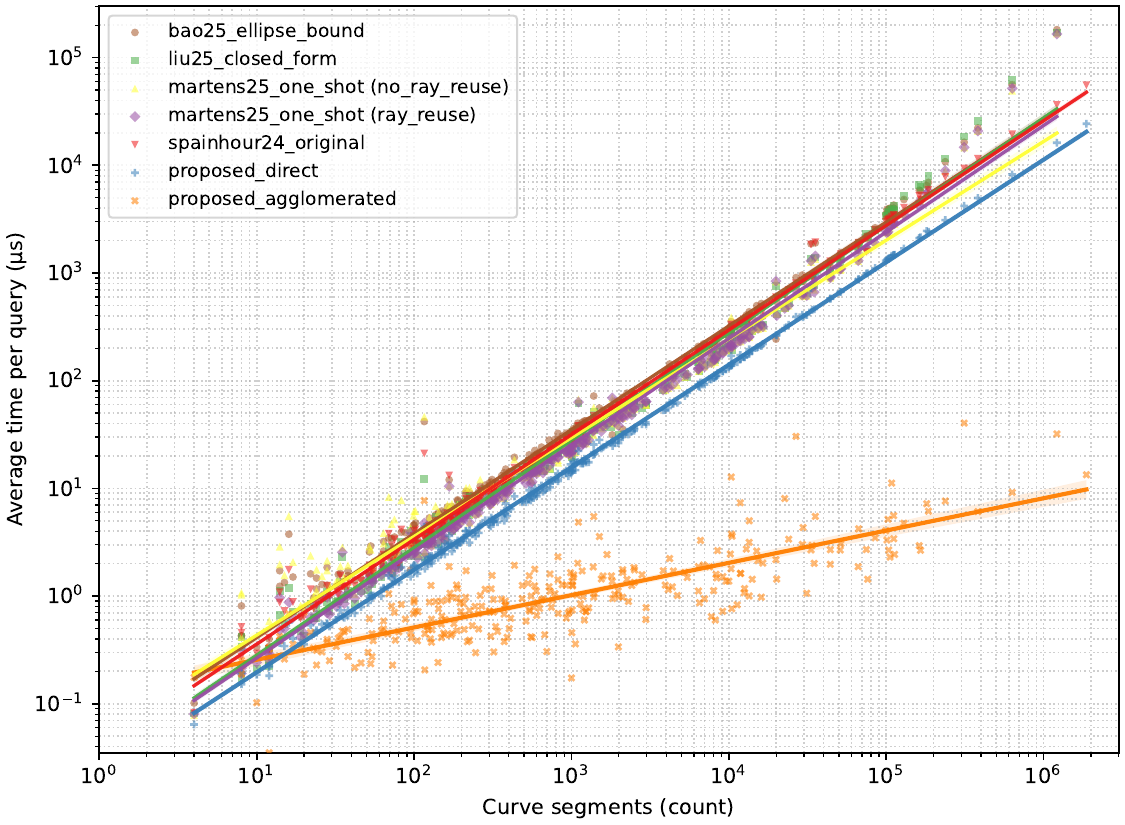}     
    \caption{Average per-query runtime (serial) for 2D direct GWN algorithms on our representative sample of 2D SVG shapes at a resolution of $500^2$.}
    \Description[Direct-method runtime comparison]{Direct-method runtime comparison}
    \label{fig:direct_experiment}
\end{figure}

We also compare our proposed 2D direct evaluation method, which has been optimized beyond the reference implementation in~\citet{spainhour_24_robustcontainment2d}, as described in Section~\ref{sec:direct_eval}.
This evaluation is performed on our representative sample of 300 SVG images from OpenClipArt20k using a $500^2$ grid of sample points. 
In addition to~\citet{spainhour_24_robustcontainment2d}, we compare our proposed direct evaluation method to~\citet{liu-2025-closedform-wn} and the preprint~\citet{bao-2025-ellipsemethod}.
We also compare to the two presented versions of~\citet{martens-2025-oneshot}, which are distinguished by whether or not cast rays are reused across query points, a strategy only available to regular grids of queries.
Because source code for these methods is currently unavailable, we compare in this experiment our best effort implementation of those methods, which we have released in an open-source repository.
To accommodate these other implementations, each method is evaluated serially.

We present this evaluation in Figure~\ref{fig:direct_experiment}, and as expected, we see that our agglomerated 2D method rapidly outpaces all direct alternatives.
We also see nearly identical scaling between each direct method, as in the absence of any spatial index, each method evaluates the GWN for most curves at a given query point with the same \texttt{atan2} call (with the exception of~\citet{spainhour_24_robustcontainment2d}, which instead utilizes \texttt{acos}).
In this way, the primary distinguishing factor between each method is how curves are processed when the query point is within their bounding box, and in this case our proposed 2D method achieves at minimum $2\times$ speedup across the entire sample.
This points to both the overall improvement of our optimized 2D method, and to its utility as the leaf-node fall back for our proposed agglomeration algorithm.

\section{Limitations \& Conclusions}
We have proposed a robust, fast GWN evaluation method for curved NURBS input in 2D and 3D. Our method evaluates the GWN on the underlying parametric representation for patches near the query point and approximates more distant geometry via a spatial index of precomputed agglomerated moment data. Our approach makes accurate point containment queries practical on large categories of in-the-wild CAD models for which direct evaluation is too slow.
Looking forward, we are interested in further optimizing our algorithm, such as through a geometry-aware definition of $\beta$, and through a GPU port.

Our aggregate testing emphasizes the importance of reliable parsing and processing of CAD objects.
For example, while per-surface parametric representation of trimming curves is the default for the STEP and IGES interchange formats, we have encountered cases in which the parsed result is malformed, resulting in surfaces with open boundaries.
These local defects propagate as their moments aggregate up the tree, which can lead to global approximation errors as incorrect values.
We are planning to investigate methods which more carefully handle trimming curves within the 3D GWN calculation to safeguard against such invalid input. 
\vspace{-.5em}

\begin{acks}
We would like to thank the anonymous reviewers for their thoughtful consideration and insightful feedback.
This work was performed under the auspices of the U.S. Department of Energy by Lawrence Livermore National Laboratory under Contract DE-AC52-07NA27344.
\end{acks}

\bibliographystyle{ACM-Reference-Format}
\bibliography{citations}

\appendix

\end{document}